\pdfoutput=1
\documentclass[journal]{IEEEtran}

\pdfpageattr{/Group << /S /Transparency /I true /CS /DeviceRGB >>}

\usepackage[utf8]{inputenc}
\usepackage[T1]{fontenc}
\usepackage{textcomp}
\usepackage{silence}
\usepackage{fixltx2e}
\usepackage{microtype}
\usepackage{calc}
\usepackage[normalem]{ulem}
\usepackage{balance}

\usepackage{lipsum}

\usepackage[range-phrase=--,per-mode=symbol-or-fraction,binary-units=true,range-units=single,list-units=single,detect-all]{siunitx}
\usepackage{silence}
\AtBeginDocument{
	\DeclareSIUnit\bit{bit}
	\DeclareSIUnit\byte{Byte}
	\DeclareSIUnit\decibeli{dBi}
	\DeclareSIUnit\decibelm{dBm}
	\DeclareSIUnit\mph{mph}
	\DeclareSIUnit\resourceblock{RB}
	\DeclareSIUnit\vehicle{veh}
	\DeclareSIUnit\watthour{Wh}
}

\usepackage{csquotes}
\usepackage[backend=biber,style=ieee,doi=false,isbn=false,mincitenames=1,maxcitenames=2]{biblatex}
\DeclareFieldFormat{sentencecase}{#1} 
\DeclareFieldFormat{titlecase}{#1} 
\usepackage{xpatch}
\xpatchbibmacro{textcite}{\addspace}{\addnbspace}{}{}
\xpatchbibmacro{Textcite}{\addspace}{\addnbspace}{}{}
\DefineBibliographyStrings{english}{
	andothers = et~al\adddot\addspace
}

\usepackage{amsmath}

\usepackage{amssymb}
\usepackage{amsfonts}
\usepackage{amsthm}

\usepackage{booktabs}
\usepackage{url}

\usepackage[inline]{enumitem}
\usepackage{subfigure}
\usepackage[pdftex]{graphicx}
\DeclareGraphicsExtensions{.pdf,.png,.jpg,.tikz}
\usepackage{tikz}
\usetikzlibrary{arrows}
\usetikzlibrary{calc}
\usetikzlibrary{chains}
\usetikzlibrary{scopes}

\usepackage[american]{babel}
\usepackage{hyphenat}

\usepackage[capitalize,noabbrev]{cleveref}
\crefformat{footnote}{#2\footnotemark[#1]#3}

\RequirePackage{xstring}
\RequirePackage{xparse}
\RequirePackage[]{acro}
\NewDocumentCommand\acrodef{mO{#1}mG{}}{\DeclareAcronym{#1}{short={#2}, long={#3}, #4}}

\usepackage{bm}
\usepackage{algorithm}
\usepackage{algpseudocode}
\usepackage{comment}
\usepackage{textcomp}
\usepackage{transparent}
\usepackage{amsmath}
\usepackage{todonotes}

\NewDocumentCommand\IEEE{ s m d[] }{%
	\IfBooleanTF{#1}{}{IEEE\,}
	\nolinebreak[2]
	#2%
	\IfNoValueTF{#3}{%
		}{%
		\StrGobbleLeft{#3}{1}[\sommerIEEEFirstLetter]%
		\IfEq{\sommerIEEEFirstLetter}{}{%
			#3
			}{%
			\nolinebreak[3]
			\StrLeft{#3}{1}%
			\sommerIEEELettersSlashed{\sommerIEEEFirstLetter}%
		}%
	}%
}
\newcommand{\sommerIEEELettersSlashed}[1]{%
	/
	\StrLeft{#1}{1}%
	\StrGobbleLeft{#1}{1}[\sommerIEEESubsequentLetter]%
	\IfEq{\sommerIEEESubsequentLetter}{}{%
		}{%
		\sommerIEEELettersSlashed{\sommerIEEESubsequentLetter}
	}%
}

\begin{document}


\title{A Novel Pulse Radar Framework for ISAC-FD-enabled URLLC Systems}

\author{%
Atefeh Rezaei, \textit{Member, IEEE},
Ata Khalili, \textit{Member, IEEE},
Falko Dressler, \textit{Fellow, IEEE},
\\
and Robert Schober, \textit{Fellow, IEEE}
\thanks{ A. Rezaei and F. Dressler are with the School of Electrical Engineering and
Computer Science, TU Berlin, Germany (e-mail: rezaei@ccs-labs.org, dressler@ccs-labs.org). }
\thanks{ A. Khalili and R. Schober are with the Institute for Digital Communications, Friedrich-Alexander-University Erlangen–Nurnberg, 91054 Erlangen, Germany (e-mail: ata.khalili@fau.de, robert.schober@fau.de).}
\thanks{This work was supported by the Federal Ministry of Research, Technology, and Space (BMFTR, Germany) within the project xG-RIC under grant 16KIS2429K as well as by the German Research Foundation (DFG) within the project RADCOM-HETNET under grant DR 639/18-4.}} 

\maketitle
\begingroup
\renewcommand{\thefootnote}{}
\footnotetext{This work has been submitted to the IEEE for possible publication. Copyright may be transferred without notice, after which this version may no longer be accessible.}
\endgroup
\begin{abstract}\nohyphens{%
We propose a novel pulse radar framework for integrated sensing and communication (ISAC) systems that supports full-duplex (FD) ultra-reliable low-latency communication (URLLC). Motivated by the stringent delay and reliability requirements of sixth-generation (6G) URLLC services, our framework integrates pulse-based radar sensing with sporadic packet transmissions for URLLC users while accounting for the effect of residual self-interference (SI) inherent in FD systems. To enable practical deployment, we adopt a time-structured approach, where the silent intervals of the radar pulse are opportunistically exploited for communication. A key focus of our work is on minimizing the total energy consumption of the base station, which is critical for sustainable radar-assisted wireless networks. To this end, we jointly optimize the sensing pulse width, beamforming vectors, and URLLC scheduling while ensuring communication and sensing quality-of-service (QoS) constraints. We formulate a non-convex mixed-integer optimization problem to address sensing and communication requirements jointly. The sensing performance, quantified via the Cramér–Rao bound (CRB), is reformulated using the Schur complement. Finite blocklength capacity approximations capture the URLLC constraint, while probabilistic constraints are modeled using cumulative distribution functions.
The original non-convex problem is first decomposed into two sub-problems via an alternating optimization algorithm.
Simulation results reveal important trade-offs between sensing time, energy consumption, and residual SI.
Our proposed method improves system performance while explicitly accounting for SI, and demonstrates improved URLLC reliability with reduced energy consumption.
The results highlight that proper design of the sensing duration and FD scheduling provides a flexible ISAC framework capable of robust operation even in strong interference conditions.
}\end{abstract}


\acresetall%
\IEEEpeerreviewmaketitle%


\section{Introduction}

Integrated sensing and communication (ISAC) is expected to play a pivotal role in future communication networks, particularly with the emergence of sixth-generation (6G) technologies.
It enables significant benefits, including improved spectral and energy efficiency, simplified hardware architectures, and reduced signaling overhead \cite{liu2022integrated}.
Furthermore, ISAC technology has the potential to enable a wide range of emerging applications such as autonomous vehicles, smart cities, and the Internet of Things (IoT) \cite{cui2021integrating,saad2020avision}.
This is made possible through accurate localization, environmental monitoring, and real-time communication updates to enhance network status awareness. Despite its numerous advantages, the practical implementation of ISAC systems introduces several challenges spanning hardware design, signal processing, and resource allocation. \cite{liu2022integrated}.
Among these challenges, resource allocation plays a particularly important role due to the strong coupling between sensing and communication functionalities under limited radio resources \cite{cui2021integrating,saad2020avision}.
In particular, transmit power allocation, beamforming design, and time scheduling must be carefully coordinated to balance sensing accuracy and communication performance.
Consequently, the joint optimization of sensing and communication resources has emerged as a fundamental problem in ISAC systems.

Beyond resource allocation, the performance of ISAC systems also strongly depends on the underlying transceiver architecture.
In particular, enabling simultaneous sensing and communication over the same frequency band requires full-duplex (FD) operation.
FD transceivers are especially attractive for ISAC since they enable concurrent data transmission and real-time monostatic sensing \cite{barneto2019full,hassani2022joint}.
However, FD operation inherently introduces strong self-interference (SI) due to the leakage of the transmitted signal into the receive chain during simultaneous transmission and reception.
Although substantial progress has been achieved in SI mitigation techniques, including passive isolation, analog cancellation, and digital suppression, complete SI elimination remains infeasible in practice.
Consequently, residual SI persists and can significantly degrade both sensing accuracy and communication reliability.

In addition to sensing accuracy and communication reliability, 6G ISAC systems are also expected to support ultra-reliable low-latency communication (URLLC) services for mission-critical applications with stringent latency and reliability requirements.
Existing studies on URLLC have investigated resource-allocation strategies to meet these requirements relying on, e.g., power control, bandwidth allocation, and short-packet transmission design \cite{popovski2019wireless,rezaei2023resource,pocovi2018achieving}.
Since achieving URLLC may require additional transmit power and communication resources, recent work has also considered energy-efficient URLLC designs that minimize power or energy consumption subject to latency and reliability constraints \cite{jalali2022powerefficient,ganjalizadeh2023saving,zou2024energyefficient}.

Along these lines, related work on ISAC, full-duplex operation, and URLLC addresses many of the identified challenges, yet looking at these aspects often in a more isolated manner.
However, existing studies rarely consider energy minimization for the overall FD-enabled ISAC system while jointly accounting for sensing scheduling, adaptive pulse-duration optimization, beamforming, residual SI, and finite-blocklength URLLC constraints.
To address these challenges, we propose a novel FD-enabled ISAC framework based on a dual-function radar communication base station (DFRC-BS).
Specifically, we develop a unified transmission structure that enables the coexistence of sensing and URLLC services while explicitly accounting for residual SI.
Based on this framework, we formulate a joint optimization problem for sensing–communication scheduling, pulse duration optimization, and transmit beamforming under finite blocklength URLLC constraints.
The proposed design captures the fundamental trade-off between sensing reliability and low-latency communication performance in FD-enabled ISAC systems.

Our key contributions can be summarized as follows:
\begin{itemize}
\item We propose a novel FD-enabled ISAC scheduling framework that leverages radar silent intervals to support URLLC transmissions. The proposed design enables the coexistence of sensing and low-latency communication within a unified frame structure. 

\item We develop a unified FD-ISAC system model that explicitly accounts for residual SI and its impact on both sensing and communication. In this context, we derive the CRB for target estimation, revealing the coupling between communication design and sensing accuracy.

\item To address energy consumption in the considered FD pulse radar ISAC system, we formulate a joint optimization problem that minimizes the total transmit energy while accounting for residual SI, radar range constraints, and per-slot power limits. Specifically, the proposed approach jointly optimizes sensing–communication scheduling over time slots, sensing pulse duration, and transmit beamforming under finite blocklength URLLC QoS constraints.

\item Simulation results demonstrate that the proposed framework substantially reduces energy consumption while preserving sensing accuracy and communication reliability, even in the presence of residual self-interference. Moreover, the proposed joint optimization framework reveals the intricate trade-off between sensing pulse duration, residual self-interference, and URLLC reliability, highlighting how adaptive pulse-width optimization enables more energy-efficient FD-ISAC operation under stringent QoS requirements.
\end{itemize}

\textit{Notations}: Diag[$\textbf{A}$] denotes a diagonal matrix with the same main diagonal elements as \textbf{A} and diag($\mathbf{a}$) denotes a diagonal matrix whose diagonal entries are given by the elements of vector $\mathbf{a}$.
$\text{Tr}(\boldsymbol{A})$ and $\boldsymbol{A}^H$ denote the trace of a matrix and the conjugate transpose of a matrix, respectively.
Furthermore, the Frobenius norm of a matrix and the Euclidean norm of a vector are denoted by $\|\boldsymbol{A}\|_F$ and $\|\boldsymbol{a}\|$, respectively.
$\text{Re}\{\cdot\}$ and  $\mathbb{E}\{\cdot\}$ are the real part and the expected value of the associated arguments, respectively.
The number of linearly independent row vectors of a matrix is denoted by rank($\cdot$), and $ \mathbf{A}\succeq \mathbf{0} $ indicates that $\textbf{A}$ is a positive semi-definite (PSD) matrix.


\section{Related Work}

Despite its importance, the coexistence of ISAC and URLLC remains relatively underexplored \cite{zhu2022power,ding2022joint,zhao2024joint,zhang2024statistical,qin2024urllc,behdad2024joint}.
Existing work mainly investigates beamforming and resource allocation for ISAC systems under latency and reliability constraints.
For example, \textcite{zhu2022power} study power allocation for ISAC systems under finite blocklength (FBL) communication constraints while guaranteeing a minimum Cramér–Rao bound (CRB) for target estimation.
In contrast, \textcite{ding2022joint} investigate beamforming optimization under communication delay and sensing quality constraints.
\textcite{zhao2024joint} employ a partially observable Markov decision process (POMDP) framework for beamforming design under periodic URLLC traffic arrivals.
Other works consider URLLC-aware ISAC from different perspectives, including age-of-information optimization \cite{zhang2024statistical} and UAV-assisted ISAC resource allocation \cite{qin2024urllc}.

The impact of residual SI is particularly critical in ISAC systems, where sensing relies on detecting weak reflected signals that are significantly attenuated by round-trip path loss \cite{barneto2021full}.
Consequently, even moderate residual SI can degrade target estimation performance.
This strong coupling between sensing and communication introduces new challenges for FD-enabled ISAC systems, since communication transmission directly affects sensing reliability through the residual SI level.
As a result, efficient resource allocation strategies that jointly optimize transmit power, beamforming, and time scheduling become essential under FD operation.
Although several works have investigated FD-enabled ISAC systems, e.g., \cite{he2023full-duplex,liyanaarachchi2021joint,he2023integrated}, resource management for such systems remains relatively underexplored, particularly under practical residual SI conditions.

Regarding energy consumption in ISAC systems, \textcite{xu2023integrated} investigate resource allocation in distributed antenna networks, where sensing and communication are performed in two separate phases.
Considering the limited capacity of the fronthaul links and the QoS requirements of both functionalities, they minimize the total energy consumption of all BSs over a given time horizon by jointly optimizing the information beamformers and the sensing signal.
Meanwhile, \textcite{behdad2024joint} study energy-efficient distributed MIMO ISAC systems and jointly design the communication and sensing functionalities to improve the overall energy efficiency.
Furthermore, \textcite{zou2024energyefficient} develop an energy-efficient beamforming framework for multi-user ISAC systems, in which the communication and sensing objectives are jointly optimized subject to transmit-power, user-SINR, and sensing-accuracy constraints.

Collectively, these studies demonstrate that substantial energy savings can be achieved through joint optimization of transmission resources, beamforming, and sensing operations while maintaining the required communication and sensing performance. However, most existing studies primarily focus on optimizing transmit-side resources, such as beamforming, power allocation, and scheduling, while treating the radar sensing pulse duration as a fixed system parameter. As a result, the potential of adaptive pulse-duration optimization to jointly balance sensing accuracy, communication reliability, and energy consumption in FD-ISAC systems remains insufficiently explored.

Despite these contributions, several important challenges remain unresolved.
Existing URLLC-ISAC frameworks mainly rely on half-duplex or orthogonal transmission schemes and do not explicitly account for FD operation and the resulting residual SI.
Furthermore, aspects such as delay-aware scheduling, adaptive pulse duration optimization, and bursty URLLC traffic variations remain insufficiently explored.
This limitation is particularly critical in FD-enabled ISAC systems, where residual SI directly affects sensing reliability due to the weak power of reflected echo signals.
Consequently, the strong coupling between sensing and communication under FD operation must be carefully considered when designing low-latency and highly reliable ISAC frameworks.
Motivated by these limitations, this work develops an FD-enabled ISAC framework that jointly optimizes sensing pulse duration, beamforming, and sensing–communication scheduling while explicitly accounting for residual SI and finite-blocklength URLLC constraints to improve energy efficiency.


\section{System Model}
    
\begin{figure}
    \centering
    \def\svgwidth{\columnwidth}
    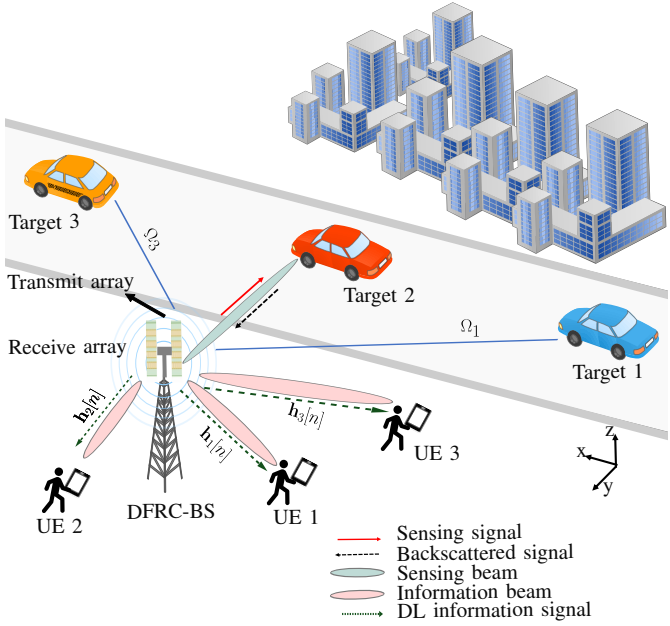
    \caption{Integrated sensing and communication in a URLLC-based network comprising multiple targets and multiple communication users.}
    \label{fig:sys}
    \vspace{-.8em}
\end{figure} 

We consider a DFRC-BS, equipped with two uniform linear arrays (ULAs), featuring $N_{t}$ transmit antennas and $N_{r}$ received antennas, respectively.
The DFRC-BS efficiently senses the environment for $E$ potential targets and serves $K$ single-antenna URLLC users, as illustrated in \Cref{fig:sys}. 
The DFRC-BS emits multiple radar pulses to sense potential targets and extracts valuable information about their characteristics, including critical network status information that enhances communication effectiveness.
We adopt a pulse radar approach to ensure the reliable detection of echoes at the transmitter.
To support real-time applications, our aim is to achieve highly reliable communication with low latency.
This is made possible by opportunistically transmitting URLLC packets during the listening periods of the radar pulses.
Therefore, we assume that the DFRC-BS can operate in FD mode, enabling effective interference cancellation between downlink data transmission and the received reflections during the listening intervals.
FD operation also reduces resource demands, such as the time allocated for sensing and communication tasks, and lowers overall energy consumption.\footnote{In our proposed system model, we assume that the transceiver can execute three stages of successive interference cancellation: passive isolation between the transmit and receive antennas, analog cancellation, and digital cancellation, as described in \cite{barneto2019full, barneto2021full}.
To fully exploit the benefits of FD-ISAC, efficient successive interference cancellation (SIC) methods are essential.
These techniques ensure the quality of the sensing and communication functionalities without imposing significant limitations on radar range or other critical parameters.}
In what follows, we provide an overview of our proposed ISAC-FD framework.

\subsection{ISAC Frame Structure}\label{isac_frame}

Our investigation focuses on how URLLC communication can be seamlessly supported while addressing the sensing task as the main functionality.
This requires that users be supported while the BS performs sensing as its main task.
When URLLC users are present, the listening time for a sensed target in time slot $n$ is concurrently utilized for downlink data transmission to the URLLC users.
The scheduler exploits the silent intervals of the sensing pulses to allocate transmission opportunities for URLLC users.
Without FD capability, the system must alternate between sensing and communication, limiting its responsiveness.
If sensing is prioritized, URLLC packets may be delayed; if URLLC is prioritized, sensing must pause, leading to reduced detection performance.
In contrast, our FD-based design pre-allocates resources to accommodate potential URLLC traffic during each frame, avoiding interruptions and ensuring that latency requirements are met without disrupting sensing tasks.

\begin{figure}
    \centering
    \def\svgwidth{\columnwidth}
    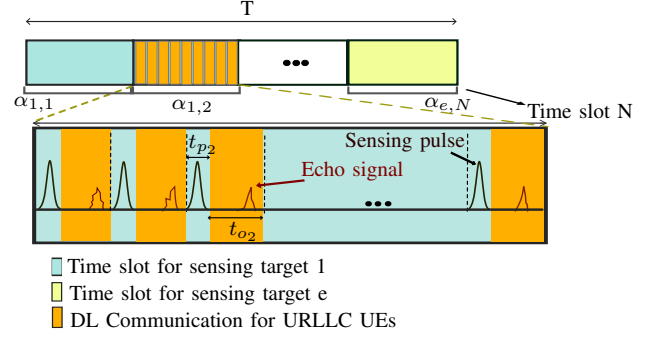
    \caption{Proposed ISAC frame structure where $T$ is the total frame duration.}%
    \label{frame}
    \vspace{-.8em}
\end{figure}

The overall frame duration $T$ is discretized into $N$ time slots, each with a duration of $\delta_t$.
\Cref{frame} illustrates the frame structure, wherein each time slot $n$ is dedicated to sensing a specific target, with the sensing covariance matrix optimized to enhance the radar’s performance.
To model this, we introduce $\alpha_{e,n}$ as the sensing indicator for target $e \in \{1,...,E\}$. If $\alpha_{e,n}=1$, it signifies the sensing of target $e$ during the $n$-th time slot; otherwise, $\alpha_{e,n}=0$.
We consider pulse-based radars that employ multiple sensing rounds to perform sensing tasks.
Hence, within each sensing round, the radar transmits a signal with a pulse width of $t_{p_n}$ and subsequently awaits the echo signal during the listening time $t_{o_n}=\dfrac{\delta_t}{N_s}-t_{p_n}$.
Importantly, we assume $N_s$ sensing rounds within time slot $n$, each with equal pulse width $t_{p_n}$.
We model user traffic as a Poisson process based on the file transfer protocol model 3 (FTP3) traffic model, which is widely used to characterize random and aperiodic traffic in wireless networks due to its analytical tractability \cite{goldsmith2005wireless,meredith2019study}. 
The URLLC users can benefit from sub-resource blocks, while the DFRC BS is listening to the sensing echoes.
As a result, we transmit one data symbol during each listening interval, $t_{o_n}$, such that a total of $N_s$ data symbols are transmitted to each user within a time slot.

\subsection{Radar Model}

The sensing signal adopted for target $e$ in time slot $n$ is denoted as $\mathbf{s}_e[n]\in {\mathbb{C}^{N_t \times 1}}$.
In the following, we describe the beampattern gain, sensing channel characterization, received echo signal, SI model, sensing range, and shape of the sensing beam in detail. 

\subsubsection{Transmit Beampattern Gain}

Radar sensing relies on the directional focusing of the emitted signal, which leads to a transmit beampattern gain directed towards the intended target.
To enable this, we assume the transmit and receive antennas at the DFRC-BS are separated to ensure sufficient isolation and effective SI suppression.
The transmit beampattern gain from the BS towards target $e$ is denoted by $\mathcal{P}(\mathbf{\hat{R}}_{e},\mathbf{\theta}_{e})[n]=\mathbf{a}_t^H(\mathbf{\theta}_{e})~ \mathbf{\hat{R}}_{e}[n]~\mathbf{a}_t(\mathbf{\theta}_{e})$ where $\mathbf{\hat{R}}_e[n]={\mathbf{s}_e[n]\mathbf{s}_e^H[n]}$.
We define $\mathbf{a}_t(\mathbf{\theta}_{e})\in {\mathbb{C}^{N_t \times 1}}$ and  $\mathbf{a}_r(\mathbf{\theta}_{e})\in {\mathbb{C}^{N_r \times 1}}$ as the steering vectors at the transmitter and receiver sides \cite{khalili2023energy-aware}:
\begin{align} \label{equ:steering}
	\hspace{-0.25mm}\mathbf{a}_{t}(\mathbf{\theta}_{e})\hspace{-0.25mm}=& \hspace{-0.25mm}\big[1,e^{j 2\pi \frac{\hat{d}}{\lambda} \sin (\theta_{e})},..., e^{j 2\pi \frac{\hat{d}}{\lambda} (N_{t}-1) \sin(\theta_{e})}\big]^T
\end{align} 
\begin{align} \label{equ:steeringr}
	\hspace{-0.25mm}\mathbf{a}_{r}(\mathbf{\theta}_{e})\hspace{-0.25mm}=& \hspace{-0.25mm}\big[1,e^{j 2\pi \frac{\hat{d}}{\lambda} \sin (\theta_{e})},..., e^{j 2\pi \frac{\hat{d}}{\lambda} (N_{r}-1) \sin(\theta_{e})}\big]^T  
\end{align} 
where $\lambda$ denotes the carrier wavelength and $\hat{d}$ represents the spacing between two adjacent transmit/receive antennas.
We consider a monostatic radar setting, in which $\mathbf{\theta}_{e}$ denotes both the direction of arrival (DoA) and the direction of departure (DoD).

\subsubsection{Received Echo Signal} 

We define the round-trip channel matrix of the $e$-th target, which is assumed to be fixed during a time slot, as  $\mathbf{H}_{e}\hspace{-1mm}=\Omega_e \mathbf{a}_{r}{(\theta_e)}\mathbf{a}_{t}^{H}{(\theta_e)} \in {\mathbb{C}^{N_r \times N_t}}$, where $\Omega_e$ denotes the complex-valued channel coefficient that dependents on the target’s radar cross section (RCS), $\nu_{\text{RCS}}$, and the round-trip path loss  as \cite{xu2023integrated}:
\begin{align} \label{equ:omega}
	\hspace{-0.25mm}\Omega_e\hspace{-0.25mm}=& \hspace{-0.25mm}\nu_{\text{RCS}}\bigg(\dfrac{c}{4\pi f_c D_e}\bigg)^2.
\end{align} 
Here, $c$ is the speed of light, $D_e$ is the distance between the radar and the $e$-th target, and $f_c$ is the carrier frequency.  

Furthermore, we assume that the same deterministic sensing signal $\mathbf{s}_e[n]$ is reused in all sensing rounds within a given time slot.
For tractability, the radar-target channel is considered quasi-static during each time slot, which is reasonable for low mobility.
In each sensing round, the radar actively transmits a pulse of short duration $t_{p_n}$. 
As a result, the echo signal received from target $e$ at the DFRC-BS in sensing round $n_{s}$, $1\leq n_{s}\leq N_{s}$, of time slot $n$ is given by:\footnote{In the proposed model of the received signal, reflections of communication signals from URLLC users to the radar receiver are neglected. This is justified by the relatively low transmit power of the communication signals compared to the power used for sensing.
Furthermore, we assume that the angular positions of the users are distinguishable from those of the targets, further supporting the assumption that user reflections can be ignored.}
\begin{align}
\mathbf{\hat{r}}_{e}[n,n_{s}]&= \alpha_{e,n}\sqrt{\frac{t_{p_n}}{\delta_t}} \mathbf{H}_{e}
\mathbf{s}_e[n] + \check{\mathbf{I}}_{\text{SI}}[n,n_s] +\check{\mathbf{z}}[n,n_s],
\end{align}
where $\check{\mathbf{z}}[n,n_{s}]$ is the received additive white Gaussian noise (AWGN) at the DFRC-BS in sensing round $n_{s}$ of time slot $n$.
The fraction $\sqrt{\frac{t_{p_n}}{\delta_t}}$ represents the proportion of time that each pulse occupies within time slot $n$.
$\check{\mathbf{I}}_{\text{SI}}[n,n_s]\in {\mathbb{C}^{N_r \times 1}}$ is the residual SI after cancellation caused by the downlink communication signal of the URLLC users in time slot $n$ and sensing round $n_s$.
The signals received during all sensing rounds within time slot $n$ are coherently combined to enhance sensing accuracy.
The resulting combined signal in time slot $n$ is given by:
\begin{align}\label{re_sens}
& \mathbf{{r}}_{e}[n]= \sum_{n_s=1}^{N_s}\mathbf{\hat{r}}_{e}[n,n_{s}]\nonumber \\ & =N_s \alpha_{e,n}\sqrt{\frac{t_{p_n}}{\delta_t}}{\mathbf{H}_{e}
\mathbf{s}_e[n]} +  {\mathbf{I}}_{\text{SI}}[n] 
+  \mathbf{z}[n].
\end{align}
where $\mathbf{z}[n]=\sum_{n_s=1}^{N_s}  \mathbf{z}[n_s,n]\sim\mathcal{C}\mathcal{N}(\mathbf{0},\sigma^{2}_{e}\mathbf{I}_{N_r})$, and  ${\mathbf{I}}_{\text{SI}}[n] =\sum_{n_s=1}^{N_s}  \check{\mathbf{I}}[n_s,n]$, which is further detailed in the following.

\subsubsection{Self-Interference} 

During the sensing listening periods, the BS transmits the information of the $k$-th URLLC user in each time slot 
$n$ using the transmit beamforming vector \( \mathbf{w}_k[n] \in \mathbb{C}^{N_t \times 1} \).
The aggregated communication signal induces SI at the radar receiver of the DFRC-BS.
Due to hardware limitations, such as the limited dynamic range of the analog-to-digital converter (ADC), residual SI persists even after cancellation.
This residual interference can be modeled as additive Gaussian noise with its covariance dependent on the communication beamforming vector.
Focusing on the dominant contribution from the SI channel, we adopt the residual SI model from \cite{day2012fullduplex}:
\begin{equation} \label{SI_first}
\mathbf{I}_{\text{SI}}[n] \sim \mathcal{CN}\left(\mathbf{0},  \mathbf{C}_{\text{SI}}[n] \right),
\end{equation}
where \( \mathbf{C}_{\text{SI}}[n]= \eta \cdot \mathrm{Diag} \left[ \mathbf{H}_{\text{SI}}  \left( N_s  \sum_{k=1}^{K} \mathbf{w}_k[n] \mathbf{w}^{H}_k[n] \right) \mathbf{H}_{\text{SI}}^{H}  \right] \). 
Here, \( \eta \) is a proportionality constant capturing the residual SI strength after suppression.
The SI channel matrix \( {\mathbf{H}}_{\text{SI}}  \in \mathbb{C}^{N_r \times N_t} \) models the coupling between the transmit and receive antennas as discussed in  \cite{he2023full-duplex}, where $[\mathbf{H}_{\text{SI}}]_{n_rn_t}= e^{-j2\pi\frac{d_{n_rn_t}}{\lambda}}$ such that  $\bar{\mathbf{H}}_{\text{SI}} = \sqrt{\eta}\mathbf{H}_{\text{SI}}$ can be considered as the effective SI channel.  
Here, $ d_{n_rn_t} $ denotes the distance between the \(n_t\)-th transmit antenna and the \(n_r\)-th receive antenna.
This residual SI contaminates the radar echo, acting as a performance bottleneck.
One of the key objectives of our framework is to ensure an accurate estimation of the target parameters despite this interference.

\begin{figure*}
\begin{equation}\label{crb}
	\zeta_{(\theta_e)}= \dfrac{ ({\sigma^2_{e}+\sigma^2_{\text{SI}}} ) \: \: \text{Tr} \big(\textbf{A}_{\theta_{e}}\mathbf{\bar{R}}_e\textbf{A}^H_{\theta_{e}}\big)}{2NN_s^2|\Omega_{e}|^2\Bigg[\text{Tr} \big(\textbf{A}^{'}_{\theta_{e}}\mathbf{\bar{R}}_e\textbf{A}^{'^H}_{\theta_{e}}\big)  \text{Tr} \big(\textbf{A}_{\theta_{e}}\mathbf{\bar{R}}_e\textbf{A}^H_{\theta_{e}}\big)-\bigg| \text{Tr} \big(\textbf{A}_{\theta_{e}}\mathbf{\bar{R}}_e\textbf{A}^{'^H}_{\theta_{e}}\big)\bigg|^2\Bigg]}.
\end{equation}
\vspace{-.8em}
\end{figure*}

\subsubsection{Sensing Range}

Following pulse radar theory, the sensing range is intricately tied to both the duration of the sensing pulse and the time allowed for the echo to arrive at the receiver \cite{skolnik2001introduction}.
Consequently, as discussed in \cref{isac_frame}, each time slot is composed of multiple sensing rounds, within each of which the BS transmits a pulse signal lasting for a duration of $t_{p_n}$.
After this emission, the BS transitions to listening mode to receive the target echo corresponding to the emitted pulse, as illustrated in \Cref{frame}.
Therefore, each sensing round operates at a specific pulse repetition frequency (PRF).
This approach allows for a meticulous balance between pulse duration and listening time, optimizing the system for accurate and reliable echo detection. 
In practical radar systems, there are limitations on the maximum and minimum detectable ranges.
Specifically, the minimum detectable distance is given by $d_{\text{min}}=\dfrac{ct_{p_n}}{2}$, and the maximum detectable distance preventing interference with other sensing rounds is $d_{\text{max}}=\dfrac{ct_{o_n}}{2}$.

\subsection{Sensing Constraint}
\label{crb_constr}

To assess the sensing accuracy, we employ the CRB for addressing the angle estimation~\cite{liu2021cramer-rao}.
However, obtaining a closed-form expression of the CRB becomes intractable in the presence of residual SI induced by FD communication \cite{liu2021cramer-rao,bekkerman2006target,he2023full-duplex,he2023integrated}.
As discussed in the previous subsection, we model the residual SI as additive zero-mean complex Gaussian noise with covariance 
\( \mathbf{C}_{\text{SI}}[n] \), which describes the distribution of interference power across the antennas at time slot $n$.
However, accounting for the instantaneous SI covariance matrix in the CRB yields an analytically intractable expression.
For analytical tractability, we approximate the residual SI as white Gaussian noise uniformly distributed across all antennas.
Specifically, we first consider the average covariance matrix over the entire time frame, yielding the time-invariant covariance matrix $\bar{\mathbf{C}}_{\text{SI}}= \eta \cdot \mathrm{Diag} \left[ \mathbf{H}_{\text{SI}}  \left( N_s \sum_{k=1}^{K}\frac{1}{N} \sum_{n=1}^N \mathbf{w}_k[n] \mathbf{w}^{H}_k[n] \right) \mathbf{H}_{\text{SI}}^{H}  \right]$.
We then approximate the full covariance matrix by a scalar variance $\sigma^2_{\text{SI}} =\mathrm{Tr} \left( \bar{\mathbf{C}}_{\text{SI}}\right)$, i.e., the trace of the diagonal elements of the averaged SI covariance matrix.
Consequently, to obtain a tractable closed-form expression for the CRB, we model the residual SI as Gaussian noise $\bar{\mathbf{I}}_{\text{SI}}[n]\sim \mathcal{CN}\left(\mathbf{0}, \sigma^2_{\text{SI}} \mathbf{I}_{N_r} \right)$, using a single time-invariant variance $\sigma^2_{\text{SI}}$. 
This approximation is consistent with evaluating the CRB over the entire frame and simplifies the incorporation of residual SI. 

Since the approximated variance $\sigma^2_{\text{SI}}$ represents the total interference accumulated across all antennas, it is generally larger than the per-antenna instantaneous SI power in \eqref{SI_first}.
As a result, designing the system to operate reliably under \(\bar{\mathbf{I}}_{\text{SI}}[n]\)  ensures robustness under the original instantaneous SI model \({\mathbf{I}}_{\text{SI}}[n]\).
Accordingly, the received signal under the described SI variance in time slot $n$ can be expressed as:
\begin{equation}\label{re_sens_wc}
\bar{\mathbf{r}}_e[n] = N_s \sqrt{\frac{t_{p_n}}{\delta_t}} \alpha_{e,n}  \, \mathbf{H}_e \mathbf{s}_e[n] + \bar{\mathbf{I}}_{\text{SI}}[n] + \mathbf{z}[n],
\end{equation}
where \( \mathbf{z}[n] \sim \mathcal{CN}(\mathbf{0}, \sigma_e^2 \mathbf{I}_{N_r}) \) is the thermal noise.
For simplicity, we define
$\beta_{e,n} = N_s\sqrt{\frac{t_{p_n}}{\delta_t}} \, \alpha_{e,n}$,
and introduce a new matrix $\mathbf{B}_e = \mathrm{diag}(\beta_{e,1}, \ldots, \beta_{e,N})$.
We define the stacked received signal matrix as:
\begin{equation}\label{Rec}
\mathbf{Y}_e =\mathbf{H}_e \mathbf{S}_e \mathbf{B}_e + \hat{\mathbf{I}}_{\text{SI}} + \mathbf{Z},
\end{equation}
where all deterministic sensing signals from different time slots are stored as \( \mathbf{S}_e = \big[\mathbf{s}_e[1], \ldots, \mathbf{s}_e[N]\big] \in \mathbb{C}^{N_t \times N} \), \( \mathbf{Y}_e = [\bar{\mathbf{r}}_e[1], \ldots, \bar{\mathbf{r}}_e[N]\big] \in \mathbb{C}^{N_r \times N} \), and \( \mathbf{Z} \) and \( \hat{\mathbf{I}}_{\text{SI}} \) are defined analogously.
Following \cite[Appendix C]{bekkerman2006target}, the CRB for estimating \( \theta_e \) for the model in \eqref{Rec} is given by \eqref{crb}, where $ \mathbf{\bar{R}}_e=\frac{1}{N}\sum_{n=1}^{N}\alpha_{e,n}\rho_n\hat{\mathbf{R}}_{e}[n]$ and $\rho_n=\dfrac{t_{p_n}} {\delta_t}$ represents the sensing duty cycle.
Matrix $\mathbf{A}_{\theta_{e}}=\mathbf{a}_{r}(\theta_{e})\mathbf{a}^{H}_{t}(\theta_{e})
$ is the product of the transmit and receive steering vectors, and $\mathbf{A}'_{\theta{e}}\in {\mathbb{C}^{N_r \times N_t}}$ represents the derivative of  $\mathbf{A}_{\theta{e}}\in {\mathbb{C}^{N_r \times N_t}}$ with respect to $\theta_e$.
This CRB expression reveals a key trade-off: the residual SI, which depends via $\sigma^2_{\text{SI}}$ on the communication beamforming vectors, directly impacts sensing accuracy.
Hence, communication and sensing designs must be jointly optimized to prevent mutual degradation.

\subsection{Communication Model}

For the data communication of URLLC users, we assume several independent signals with a duration of \(t_{o_n}\) to be transmitted in different listening times.
This means that the BS transmits one data symbol per listening interval with information signal ${c}_k[n_s,n]$, $c_k \sim \mathcal{CN} (0,1), k\in \{1, ..., K\}$.
Then, the received signal at user $k$ can be written as
\begin{equation}\label{c_signal}
{y}_k[n_s,n]=\mathbf{h}_k^{H}[n]\bigg ( \sum_{k=1}^{K} \mathbf{w}_k[n]{c}_k[n_s,n]\bigg)+n_k[n_s,n],
\end{equation}
where the channel vector \( \mathbf{h}_k[n] \in \mathbb{C}^{N_t \times 1} \) between the BS and user $k$ follows a Rician fading model and  \( {n}_k[n_s,n] \sim \mathcal{CN}({0}, \sigma^2_k) \) is the AWGN noise with variance \(\sigma^2_k\).
We assume a minimum acceptable listening time duration that guarantees sufficient communication time such that \(B\geq \frac{1}{t_{o_n}}\), where  \(B\) denotes the fixed bandwidth.
Hence, the downlink SINR for user \( k \) in sensing round \( n_s \) in time slot \( n \) is expressed as:
\begin{equation}\label{sinr}
	\tilde{\gamma}_k[n_s,n] = \frac{\big|\mathbf{h}_k^{H}[n] \mathbf{w}_k[n]\big|^2}{\sum_{i \neq k} \big|\mathbf{h}_k^{H}[n] \mathbf{w}_i[n]\big|^2 + \sigma_k^2},
\end{equation}
where \( \mathbf{h}_k[n] \in \mathbb{C}^{N_t \times 1} \) follows a Rician fading model.
Since the channel is assumed to remain constant throughout all sensing rounds within a given time slot, the transmit beamforming vector \(\mathbf{w}_k[n] \), which is the beamforming in the time slot $n$, does not change the sensing rounds of a time slot. 
As a consequence, the SINR remains unchanged across all sensing rounds within a time slot \(\gamma_k[n]=\tilde{\gamma}_k[n_s,n], \forall n_s\).

\subsection{URLLC Constraint}
\label{urllc_constr}

In URLLC systems, data packets are short and the reliability constraint is stringent.
Therefore, the Shannon capacity is not a suitable metric. Instead, we adopt the finite blocklength framework  \cite{alsenwi2021intelligent}, which  characterizes the maximum number of transmitted bits over short durations. 
In particular, the total number of bits for downlink transmission under the finite blocklength regime is given by  \cite{jalali2022powerefficient, darabi2023active,polyanskiy2010channel,ghanem2020resource}:
\begin{align}
    \label{rate}
	& r_k[n] = \nonumber \\
   &\sum_{n_s=1}^{N_s}  \log (1+\tilde{\gamma}_k[n_s,n])-Q^{-1}(\epsilon_k) \sqrt{ \sum_{n_s=1}^{N_s}\tilde{V}_k[n_s,n] }  \nonumber \\
   & =N_s\log(1 + \gamma_k[n]) - Q^{-1}(\epsilon_k) \sqrt{ N_sV_k[n] },
\end{align}
where \( \epsilon_k \) is the decoding error probability (DEP), 
\( V_k[n]=\tilde{V}_k[n_s,n] = a^2 (1 - (1 + \gamma_k[n])^{-2}) \) is the channel dispersion, with \( a = \log_2(e) \), and \( Q^{-1}(\cdot) \) is the inverse of the Gaussian Q-function.\footnote{$ Q(x)=  \int_{x}^{\infty} \frac{1}{\sqrt{2\pi}}\text{exp} (-\frac{t^2}{2}) \,dt  $} 
The total number of transmitted symbols per user during a time slot $n$ equals \(N_s\).

To ensure predictable performance, the BS proactively reserves transmission time for each user in advance.
This scheduling is guided by predicted URLLC traffic patterns.
We model the traffic load of user \( k \) in slot \( n \) as a random variable $L_{k,n} = \nu_{k,n} \tilde{\lambda}_{k,n}$, where \( \nu_{k,n} \) is the packet size (in bits) and \( \tilde{\lambda}_{k,n} \) is the packet arrival rate, modeled as a Poisson random variable.
To ensure URLLC quality-of-service (QoS) requirements, we impose a per-user probabilistic constraint on the traffic load, ensuring that the scheduled data transmission is sufficient to support the stochastic traffic load at least with probability  \( 1 - \epsilon_u \), where \( \epsilon_u \) denotes the maximum tolerable violation probability \cite{jalali2022powerefficient}. This leads to the following constraint: 
\begin{equation}\label{pr-rate}
	\Pr \big \{ L_{k,n} \geq r_k[n] \big  \} \leq \epsilon_u. \quad 
\end{equation}

\textit{Remark:}   
The system predicts the potential arrival of URLLC packets in the upcoming time slot by utilizing traffic models and QoS constraints to estimate the required resources.
A portion of the available resources is proactively reserved to ensure that, if URLLC traffic arrives, it can be served without delay.
Downlink transmissions for already-connected users are scheduled at the beginning of each time slot.
In this context, the system must proactively allocate resources for URLLC users who may need to receive their data packets in the following frame.


\section{Problem Formulation}
We formulate an optimization problem for the proposed ISAC framework, incorporating the URLLC reliability and sensing accuracy constraints introduced in the preceding subsections.
As a novel contribution, the sensing requirement is enforced via the constraint $\zeta_{(\theta_e)} \leq \zeta_{\text{max}}$, which bounds the CRB to guarantee the desired estimation accuracy.
We formulate the problem to minimize the total system energy consumption by jointly optimizing the beamforming vectors $\mathbf{w}_{k}[n]$, sensing covariance matrix $\hat{\mathbf{R}}_{e}[n]$, sensing time allocation $\alpha_{e,n}$, and pulse duration $t_{p_n}$, subject to the URLLC and sensing constraints.

Formally, the optimization problem is expressed as:
\begin{align}
\mathcal{P}_{1}:& \mathop {\rm{min}} \limits_{\boldsymbol{\Xi}} \mathcal{O}bj_1 \triangleq 
\sum_{n=1}^{N} \bigg( (\delta_t-N_st_{p_n}) \sum_{k=1}^{K} \|\mathbf{w}_k[n]\|^2 \nonumber \\ 
& \quad \quad \quad \quad \:+ N_{s} t_{p{_n}} \sum_{e=1}^{E} \alpha_{e,n}  \text{Tr}(\hat{\mathbf{R}}_{e}[n])\bigg)\nonumber\\
	\text{s.t.} ~~
	& \text{C}1:\text{Pr}\bigg \{ L_{k,n} \geq   r_k[n]\bigg \}\leq \epsilon_u, \forall k, n,\nonumber\\
&\text{C}2:\sum_{e=1}^{E}\alpha_{e,n}\leq 1, \forall n,~\nonumber\\
&\text{C}3: {\zeta_{(\theta_e)}}\leq \zeta_{\text{max}}, \: \forall e, ~\nonumber\\
&\text{C}4:\alpha_{e,n} \in \{0,1\}, \forall e,n, \nonumber\\
&\text{C}5: t_{\text{min}}\leq t_{p_n}\leq t_{\text{max}}, \forall n, \nonumber\\
&\text{C}6: d_{\text{min}} \leq D_e \leq d_{\text{max}}, \forall e, \nonumber\\
&\text{C7}:\sum_{k=1}^{K} \|\mathbf{w}_k[n]\|^2\leq p^{\text{Com}}_{\text{max}}, \forall n, \nonumber\\
 &\text{C8}:  \text{Tr}(\hat{\mathbf{R}}_{e}[n]) \leq p^{\text{Rad}}_{\text{max}}, \forall e,n.
\end{align}

In $ \mathcal{P}_{1}$, $\boldsymbol{\Xi}=\{ \mathbf{w}_{k}[n], \hat{\mathbf{R}}_e[n], {t}_{p_n}, {\alpha_{e,n}}\}$ is the set of optimization variables.
\text{C}1 ensures that the traffic load does not exceed the allocated resources with high probability. 
\text{C}2 limits the sensing process to one target per time slot, preventing overlapping target sensing.
$\text{C}3$ imposes a constraint on the CRB, ensuring adequate sensing performance. 
\text{C}4 ensures that the sensing indicator is a binary variable.
\text{C}5 establishes a minimum duration for the sensing pulse ($t_{\text{min}}$) to account for hardware limitations; thus, it imposes the upper bound (\(t_{\text{max}}\)) for the sensing pulse duration to ensure a sufficiently large listening interval for communication transmission, which leads to guarantees sufficient communication time.
\text{C}6 limits the radar sensing range to avoid interference between consecutive sensing rounds. It depends on the distance between the BS and the target $D_e$ as well as the sensing pulse duration $t_{p_n}$. 
\text{C}7 and \text{C}8 limit the power budget allocated to sensing and communication, respectively.

Optimization problem $\mathcal{P}_{1}$ is intractable and non-convex due to the multiplication of several variables and the non-convex objective function.
Additionally, the non-convex constraints C1, C3, and C4 further complicate the problem, particularly the binary variable ${\alpha}_{e,n}$  transforms the optimization into a MINLP problem. 
To address this intractability with low complexity, we propose an iterative algorithm based on the alternating optimization (AO) approach.
Specifically, we first optimize the beamforming matrices and the sensing indicator while treating  ${t}_{p_n}$ as a fixed predetermined value.
Then, we jointly optimize the pulse width and the sensing indicator based on the solutions $\mathbf{w}^*_k[n]$ and $\hat{\mathbf{R}}^*_e[n]$ obtained in the first step. 
This iterative process continues until the problem converges to a feasible solution for $\{\mathbf{w}^*_k[n], \hat{\mathbf{R}}^*_e[n], {\alpha}^*_{e,n},  {t}_{p_n}^*\}$.

\subsection{Beamforming and Sensing Indicator Optimization}

We define the product of two variables, $\alpha_{e,n}$ and $\hat{\mathbf{R}}_e[n]$, as a new auxiliary variable ${\mathbf{R}}_e[n]$, such that ${\mathbf{R}}_e[n]=\alpha_{e,n}\hat{\mathbf{R}}_e[n]$.
Besides, we employ the big-M method that facilitates the decomposition of the multiplicative terms \cite{rezaei2022energ-yefficient}.
Hence, ${\mathbf{R}}_e[n]$ is decomposable via considering the following additional constraints:
\begin{subequations} 
\begin{align}
& \Dot{\text{C}}8: \mathbf{0} \preceq {\mathbf{R}}_e[n]\preceq \alpha_{e,n}p_{\text{max}}^{\text{Rad}}\mathbf{I}\label{c-bm-1},  \\
& \Ddot{\text{C}}8 : \hat{\mathbf{R}}_e[n] -(1-\alpha_{e, n})p_{\text{max}}^{\text{Rad}} \mathbf{I}
\preceq {\mathbf{R}}_e[n]\preceq  \hat{\mathbf{R}}_e[n]. \label{c-bm-2} 
\end{align}
\end{subequations} 

C4 is a non-convex binary constraint.
To address this, we relax $ \alpha_{e, n} $ into a continuous variable.
However, to enforce approximate binary values, we introduce the following equivalent constraints, as suggested in  \cite{khalili2023energy-aware}, which guide the optimization problem to select values close to binary:
\begin{subequations} 
\begin{align}
& \Dot{\text{C}}4 : \check{Q}({\alpha}_{e, n})-E({\alpha}_{e, n})\leq 0 \label{b-1}, \\
& \Ddot{\text{C}}4 : 0 \leq {\alpha}_{e, n}\leq 1, \forall n  \in \mathcal{N},  \forall  e\in {E},  \label{b-2} 
\end{align}
\end{subequations}
where $  \check{Q}({\alpha}_{e, n})=\sum_{n\in \mathcal{N}}\sum_{e\in {E}}  \alpha_{e, n}$ and $E(\alpha_{e, n})=\sum_{n\in \mathcal{N}}\sum_{e\in {E}}  ({\alpha}_{e, n})^2$.
Note that constraint $\Dot{\text{C}}4$ is non-convex, and we replace it by considering the Taylor approximation as:
\begin{align}
& \dddot{\text{C}}4 :  \check{Q}({\alpha}_{e, n}) \nonumber \\ &- \sum_{n\in \mathcal{N}}\sum_{e\in {E}}
(\alpha^2_{e,n})^{[j]}+
2\alpha_{e, n}^{[j]}(\alpha_{e, n}-\alpha_{e, n}^{[j]})
\leq 0, 
\end{align}
where $j$ denotes the $j$-th iteration of the proposed AO approach.
Now, we introduce a penalty factor $\lambda_{1}$ to move constraint $\dddot{\text{C}}4$ to the objective function.
$\lambda_{1}$ represents the relative importance of recovering binary values for $\alpha_{e,n}$.
Hence, problem $\mathcal{P}_{1}$ can be rewritten as:
\begin{align} 
\mathcal{P}_{2}&:  \: \underset{\boldsymbol{\bar{\Xi}}}{\text{min}} \: \mathcal{O}bj_2 \triangleq \sum_{n=1}^{N} \bigg(  \sum_{k=1}^{K} (\delta_t-N_st_{p_n})  \|\mathbf{w}_k[n]\|^2 \nonumber \\ & + \sum_{e=1}^{E} N_st_{p_n}\text{Tr}({\mathbf{R}}_e[n])+\lambda_{1}\bigg(\check{Q}({\alpha}_{e, n})\nonumber \\ &- \sum_{n\in \mathcal{N}}\sum_{e\in {E}}
(\alpha^2_{e,n})^{[j]}+
2\alpha_{e, n}^{[j]}(\alpha_{e, n}-\alpha_{e, n}^{[j]})\bigg) \nonumber, \label{erlx} \\
s.t.&: \text{C1}-\text{C3},  \:  \Ddot{\text{C}}4, \: \text{C7}, \:\Dot{\text{C}}8, \text{and} \:\Ddot{\text{C}}8.  
\end{align}
 
The set of optimization variables is $\boldsymbol{\bar{\Xi}}=\{ \mathbf{w}_{k}[n], \hat{\mathbf{R}}_e[n],  {\alpha_{e,n}},{\mathbf{R}}_e[n]\}$, and constraints C1 and C3 are non-convex. 
To address the non-convexity of C1, we introduce the slack variable $\mu_{k,n}$, which characterizes the lower bound of the SINR function ($\gamma_{k}[n]$).
Consequently, we define the following additional constraint:
\begin{equation}\label{211c}
\dfrac{  \big|\mathbf{h}^{H}_k[n]\mathbf{w}_k[n]\big|^2}{\sum_{i\neq k}  \big|\mathbf{h}^{H}_k[n]\mathbf{w}_{i}[n]\big|^2 + \sigma^2_k} \geq \mu_{k,n}.
\end{equation}
Note that \eqref{211c} is non-convex.
Hence, we solve it by considering an upper bound for the denominator, denoted as $\iota_{k,n}$.
We replace \eqref{211c} with the following constraints:
\begin{align}  
& \Dot{\text{C}}1: \sum_{i\neq k}^{}\big|\mathbf{h}^H_k[n]\mathbf{w}_{i}[n]\big|^2+\sigma^2_k \leq \iota_{k,n},  \\
& \Ddot{\text{C}}1: \big|\mathbf{h}^{H}_k[n]\mathbf{w}_k[n]\big|^2 \geq \widetilde{\mu_{k,n}\iota_{k,n}}, \\
& \dddot{\text{C}}1: \mu_{k,n}\geq 0.\label{20c}
\end{align}
where $\widetilde{\mu_{k,n}\iota_{k,n}}$ is obtained utilizing the SCA approach as:  
\begin{align}
&{\mu_{k,n}\iota_{k,n}}\approx \widetilde{\mu_{k,n}\iota_{k,n}} = \frac{1}{2}(\iota_{k,n}+\mu_{k,n})^{2}\nonumber\\& -\frac{1}{2}\big((\iota_{k,n}^{2})^{[j]}+(\mu_{k,n}^{2})^{[j]}\big)-\iota_{k,n}^{[j]}(\iota_{k,n}-\iota_{k,n}^{[j]})\nonumber\\&-\mu_{k,n}^{[j]}(\mu_{k,n}-\mu_{k,n}^{[j]}).
\end{align}
where $\iota_{k,n}^{[j]}$, and $\mu_{k,n}^{[j]}$ are the solutions in the $j$-th iteration.
Since $\Ddot{\text{C}}1$ still is not convex, we consider the following surrogate function in which we provide an affine function for the right hand side of the constraint $\Ddot{\text{C}}1$ via Taylor approximation as:    
\begin{align}
&\big|\mathbf{h}^{H}_k[n]\mathbf{w}_k[n]\big|^2\approx \nonumber \\  & 2\mathcal{R}\bigg( (\mathbf{h}_{k}^H[n] {\mathbf{w}^{[j]}_k[n]})^{H}\mathbf{h}_{k}^H[n] \mathbf{w}_k[n]\bigg)-{\left|\mathbf{h}_{k}^H[n] {\mathbf{w}^{[j]}_k[n]} \right|}^2.
\end{align}
C1 is a probabilistic constraint, which makes it intractable to solve directly.
Consequently, constraint C1 is transformed into the following equivalent constraint:
\begin{align}
\ddddot{\text{C}}_1:  r_k[n] \geq \nu_{k,n}F^{-1}_{{\tilde{\lambda}}_{k,n}}(1-\epsilon_u), \forall k, n,
\end{align}
where $F^{-1}_{{\tilde{\lambda}}_{k,n}}$ is the inverse of the cumulative distribution function (CDF) of the packet arrival rate of the $k$-th URLLC user.
C1 is still non-convex.
Hence, we introduce an approximation for the rate function as:
\begin{align}
  \tilde{r}_{k}[n]\approx  F_{k,n}-\tilde{G}_{k,n},
\end{align}
where $F_{k,n}=N_s\text{log}(1+\mu_{k,n})$ and 
\begin{align}
\tilde{G}_{k,n}=Q^{-1}_{(\epsilon_k)}\sqrt{N_sa^2}
\bigg(  
\sqrt{1-\frac{1}{(1-\mu_{k,n}^{[j]})^{2}}}+\nonumber\\\big((1+\mu_{k,n}^{[j]})^{3}\sqrt{\frac{\mu_{k,n}^{[j]}(\mu_{k,n}^{[j]}+2)}{(1+\mu_{k,n}^{[j]})^2}}\big)^{-1}(\mu_{k,n}-\mu_{k,n}^{[j]})
\bigg).
\end{align}
By imposing the obtained approximation of $r_k[n]$ on $\ddddot{\text{C}}_1$, we obtain the following convex constraint:
\begin{align}
\hat{\text{C}}_1:  F_{k,n}-\tilde{G}_{k,n}  \geq \nu_{k,n}F^{-1}_{{\tilde{\lambda}}_{k,n}}(1-\epsilon_u).
\end{align}

Constraint C3 is highly non-linear and non-convex, as it involves the multiplication of several variables in a fractional function. 
We note that $\rho$ is considered fixed during the beamforming optimization, according to sub-problem $\mathcal{P}_2$. 
To solve C3, we introduce a new slack variable $t_e$, which allows us to restate constraint C3 using the Schur complement condition, resulting in $\Dot{\text{C}}3$ and $\Ddot{\text{C}}3$ where $\tilde{\textbf{R}}_e=\frac{1}{N}\sum_{n=1}^{N}\rho_n{\mathbf{R}}_e[n]$ because the big-M approach was employed:
\begin{subequations} \label{crb_f}
\begin{align}
& \Dot{\text{C}}3: \:   \dfrac{ \sigma_{\text{SI}}^2+\sigma_e^2 }{2|\Omega_e|^2NN_s}\leq \zeta_\text{max}t_e, \forall e,\\
& \Ddot{\text{C}}3: \: {\begin{bmatrix}
\text{Tr}\big(\textbf{A}^{'}_{\theta_e}\tilde{\textbf{R}}_e\textbf{A}^{'^H}_{\theta_e}\big)-t_e &\text{Tr}\big(\textbf{A}_{\theta_e} \tilde{\textbf{R}}_e\textbf{A}^{'^H}_{\theta_e}\big )\\
\text{Tr}\big(\textbf{A}_{\theta_e} \tilde{\textbf{R}}_e\textbf{A}^{'^H}_{\theta_e}\big ) &  \text{Tr}\big(\textbf{A}_{\theta_e} \tilde{\textbf{R}}_e\textbf{A}^H_{\theta_e}\big )
\end{bmatrix}}\succcurlyeq \textbf{0}, \forall e. 
\end{align}
\end{subequations}
Consequently, the beamforming problem is transformed into the following convex optimization problem, which can be efficiently solved using convex optimization tools:
\begin{align}  \label{relaxed-1} 
\mathcal{P}_{3}&:  \: \underset{\boldsymbol{\tilde{\Xi}}}{\text{min}} \: \mathcal{O}bj_2 \nonumber\\
s.t.&: \hat{\text{C}}1, \: \Dot{\text{C}}1,  \:\Ddot{\text{C}}1, \: \dddot{\text{C}}1, \text{C}2, \:  \dot{\text{C}}3, \:  \Ddot{\text{C}}3,  \: \Ddot{\text{C}}4,\nonumber\\ &  \text{C}7 ,\:\Dot{\text{C}}8, \: \text{and}  \:\Ddot{\text{C}}8.  
\end{align}
 
In $\mathcal{P}_{3} $, we define the optimization variable set as $\boldsymbol{\tilde{\Xi}}=\{ \mathbf{w}_{k}[n], \hat{\mathbf{R}}_e[n], {\mathbf{R}}_e[n], {\alpha_{e,n}}, {\mu_{k,n}}, {\iota_{k,n}}, t_e\}$

\subsection{Radar Pulse Width Optimization}

We design the radar pulse, utilizing the obtained solutions for $\{\mathbf{w}^*_{k,n}, \hat{\mathbf{R}}^*_e[n]\}$. 
As a result, $\mathcal{P}_1$ is transformed into the following sub-problem:
\begin{align}\label{relaxed-2} 
\mathcal{P}_{4}&:  \: \underset{\boldsymbol{\check{\Xi}}} {\text{min}} \: \mathcal{O}bj_1 \nonumber\\
s.t.&:  \text{C}2, \: \text{C}3,  \: \text{C}4, \: \text{C}5, \: \text{and}\: \text{C}6. 
\end{align}
In $\mathcal{P}_4$, constraints C3 and C4 contain non-convex functions involving the variable set $\boldsymbol{\check{\Xi}}=\{ \alpha_{e,n}, t_{p_n}\}$. 
Due to the multiplication of variables $t_{p_n}$ and $\alpha_{e,n}$, we apply the big-M method by introducing a new auxiliary variable $\tilde{t}_{p_{e,n}}=\alpha_{e,n}{t}_{p_n}$. 
This facilitates the decomposition of the term.
Therefore, the following constraint must hold to ensure the valid decomposition of $\tilde{t}_{p_{e,n}}$:
\begin{subequations} 
\begin{align}
& \text{C}9: 0 \leq \tilde{t}_{p_{e,n}}\leq \alpha_{e,n}t_{\text{max}}\label{c-bm-t1},  \\
& \Dot{\text{C}}9 : {t}_{p_n} -(1-\alpha_{e, n})t_{\text{max}} 
\leq \tilde{t}_{p_{e,n}}\leq  {t}_{p_n}, \label{c-bm-t2} 
\end{align}
\end{subequations} 
in which $t_{\text{max}}$ is an upper bound for ${t}_{p_n}$.
Constraint $\text{C}3$, which involves $\alpha_{e,n}$ and  $t_{p_{n}}$, is non-convex. 
To address this, we apply the same strategy based on the employed Schur complement condition.
Using the big-M method, we define new variables such that $\tilde{t}_{p_{e,n}}={t}_{p_{e,n}} \alpha_{e,n}$ and $\tilde{\rho}_n = \dfrac{\tilde{t}_{p_{e,n}}}{\delta_t}$, resulting in the constraint in \eqref{crb_f_1}, where $\check{\textbf{R}}_e= \frac{1}{N}\sum_{n=1}^{N}\tilde{\rho}_n \hat{\mathbf{R}}_e[n]$:
%
\begin{subequations} \label{crb_f_1}
\begin{align}
& \tilde{\text{C}}3: \: {\begin{bmatrix}
\text{Tr}\big(\textbf{A}^{'}_{\theta_e}\check{\textbf{R}}_e\textbf{A}^{'^H}_{\theta_e}\big)-t_e &\text{Tr}\big(\textbf{A}_{\theta_e} \check{\textbf{R}}_e\textbf{A}^{'^H}_{\theta_e}\big )\\
\text{Tr}\big(\textbf{A}_{\theta_e} \check{\textbf{R}}_e\textbf{A}^{'^H}_{\theta_e}\big ) &  \text{Tr}\big(\textbf{A}_{\theta_e} \check{\textbf{R}}_e\textbf{A}^H_{\theta_e}\big )
\end{bmatrix}}\succcurlyeq \textbf{0}, \forall e.\nonumber
\end{align}
\end{subequations}

As observed in the first sub-problem, the binary variable $\alpha_{e,n}$ in C4 can be relaxed and replaced with $\Ddot{\text{C}}4$ and $\dddot{\text{C}}4$.
Next, we incorporate constraint $\dddot{\text{C}}4$, which is convex due to the Taylor approximation, together with the penalty factor $\lambda_1$.
Since  $\dot{\text{C}}3$ and $\tilde{\text{C}}3$ are convex, this leads to the following convex optimization problem: 
\begin{align}\label{relaxed-21k}
\mathcal{P}_{5}&:  \: \underset{ \boldsymbol{\hat{\Xi}}}{\text{min}} \:\mathcal{O}bj_3\triangleq \mathcal{O}bj_1  + \lambda_{1}\bigg(\check{Q}({\alpha}_{e, n})\nonumber\\ &- \sum_{n\in \mathcal{N}}\sum_{e\in {E}}
(\alpha^2_{e,n})^{[j]}+
2\alpha_{e, n}^{[j]}(\alpha_{e, n}-\alpha_{e, n}^{[j]})\bigg)\nonumber\\
s.t.&:\nonumber\\ &  \text{C}2,\: \tilde{\text{C}}3,\:\dot{\text{C}}3, \:\Ddot{\text{C}}4, \text{C}5,\:  \text{C}6, \: \text{C}9,\: \Dot{\text{C}}9, 
\end{align}
where the variable set is $\boldsymbol{\hat{\Xi}}=\{ \alpha_{e,n},t_{p_n}, \tilde{t}_{p_n}, t_e\}$.

\subsection{Overall Algorithm}

\begin{algorithm}[t]
\caption{Proposed resource allocation framework}
\label{alg1}
\renewcommand{\arraystretch}{1.2}

\textbf{Input:} Select feasible values for initializing the parameters:
$\mathbf{w}^{[j]}_{k}[n]$, $\hat{\mathbf{R}}^{[j]}_e[n]$, $\alpha_{e,n}^{[j]}$,
$t_{p_n}^{[j]}$, iteration index $j$, and $\epsilon$.


\textbf{Repeat:}
\begin{enumerate}
    \item Solve $\mathcal{P}_3$ to obtain
    $\mathbf{w}^{[j+1]}_{k}[n]$, $\hat{\mathbf{R}}^{[j+1]}_e[n]$, $\alpha_{e,n}^{[j+1]}$
    based on $t = t_{p_n}^{[j]}$.

    \item Solve $\mathcal{P}_5$ to obtain
    $\alpha_{e,n}^{[j+1]}$ and $t_{p_n}^{[j+1]}$
    based on $\mathbf{w}_{k}[n] = \mathbf{w}^{[j+1]}_{k}[n]$,
    $\hat{\mathbf{R}}_e[n] = \hat{\mathbf{R}}^{[j+1]}_e[n]$.

    \item Update iteration index: $j \leftarrow j + 1$.

    \item Check convergence:
    \[
    \frac{\mathcal{O}bj^{[j]} - \mathcal{O}bj^{[j-1]}}{\mathcal{O}bj^{[j-1]}} \leq \epsilon.
    \]
\end{enumerate}


\textbf{Output:}
$(\mathbf{w}^*_{k}[n], \hat{\mathbf{R}}^*_e[n], \alpha_{e,n}^*, t_{p_n}^*)$.

\end{algorithm}

\Cref{alg1} outlines the AO-based solution.
In the following, we discuss its convergence properties and computational complexity.

\subsubsection{Convergence}
The proposed AO algorithm alternately refines two variable groups:
(i) $\boldsymbol{\Pi} \triangleq \{ \hat{\mathbf{R}}_e[n], \mathbf{w}_{k}[n], \alpha_{e,n}\}$, which includes the beamforming vectors for sensing and communication as well as the sensing indicators, and
(ii) $\boldsymbol{\bar{\Pi}} \triangleq \{ \alpha_{e,n},  t_{p_n} \}$, associated with sensing indicator and pulse width parameters.
The algorithm leverages SCA to address the non-convex nature of the optimization problem.
When optimizing $\boldsymbol{\Pi}$ with fixed $t_{p_n}$, problem $\mathcal{P}_3$ is solved such that the objective satisfies
$\mathcal{O}bj_2(\boldsymbol{\Pi}^{[j+1]},  t_{p_n}^{[j]}) \leq \mathcal{O}bj_2(\boldsymbol{\Pi}^{[j]},  t_{p_n}^{[j]})$.
Similarly, updating $\boldsymbol{\bar{\Pi}}$ while holding $\mathbf{w}_{k}^{[j+1]}[n]$ and $\hat{\mathbf{R}}^{[j+1]}_e[n]$ constant yields $\mathcal{O}bj_2(\boldsymbol{\Pi}^{[j+1]},  t_{p_n}^{[j+1]}) \leq \mathcal{O}bj_2(\boldsymbol{\Pi}^{[j+1]},  t_{p_n}^{[j]})$.
This guarantees that the objective function in $\mathcal{P}_1$ is non-increasing or remains unchanged with each iteration of \cref{alg1}, resulting in a suboptimal solution  \cite{xu2023integrated,khalili2025movable-preprint,tseng2001convergence}.

\subsubsection{Computational Complexity}
Next, we analyze the computational complexity based on the SCA method described in \cref{alg1}.
The complexity is given by $\mathcal{O}\Big(\mathrm{log}(1/\epsilon)m_{\text{PS}}n_{\text{CS}}^3 \Big)$, where $\mathcal{O}(\cdot)$ denotes the big-O notation and $\epsilon$ represents the desired solution accuracy \cite{boyd2004convex}. 
Here, $m_{\text{PS}}$ denotes the problem size, and $n_{\text{CS}} $ is the number of constraints.
For problem $\mathcal{P}_{3}$, the problem size and number of constraints are given by $m_{\text{PS}}= KN+EN$ and $n_{\text{CS}}= 4KN+ 4EN+ N+2E$.
For problem $\mathcal{P}_{5}$,  they are $m_{\text{PS}}= EN$ and $n_{\text{CS}}= 5EN+N+3E$.


\section{Simulation Results}

In the following, we evaluate the effectiveness of the proposed algorithm in minimizing energy consumption as the primary objective, while satisfying sensing time requirements in ISAC-enabled systems.
Furthermore, we analyze the sensitivity of the algorithm with respect to different bounds and key parameters in the CRB and URLLC constraints, demonstrating that stricter requirements can be achieved at the expense of increased energy and time consumption.
In addition, we assess the proposed sensing time scheduling under various system configurations and compare it with benchmark schemes, including scenarios without pulse width optimization and without URLLC constraints.
\begin{figure*}
\centering
\subfigure[Total energy consumption vs.\ $\eta$.]{
\label{E_SI_a}
\includegraphics[width=0.475\textwidth]{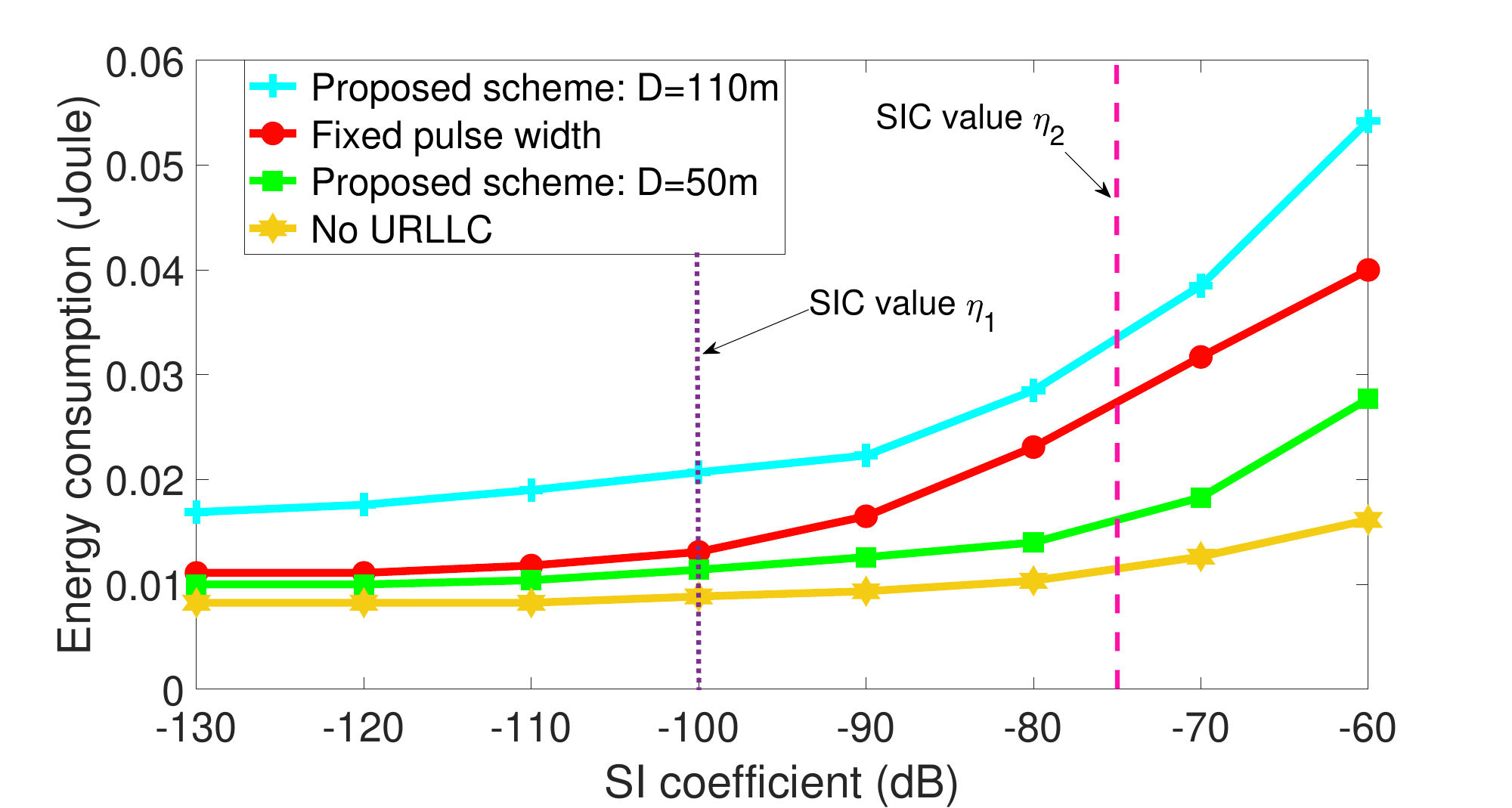}
}
\hfill
\subfigure[Total required active sensing time vs.\ $\eta$.]{
\label{S_SI_a}
\includegraphics[width=0.475\textwidth]{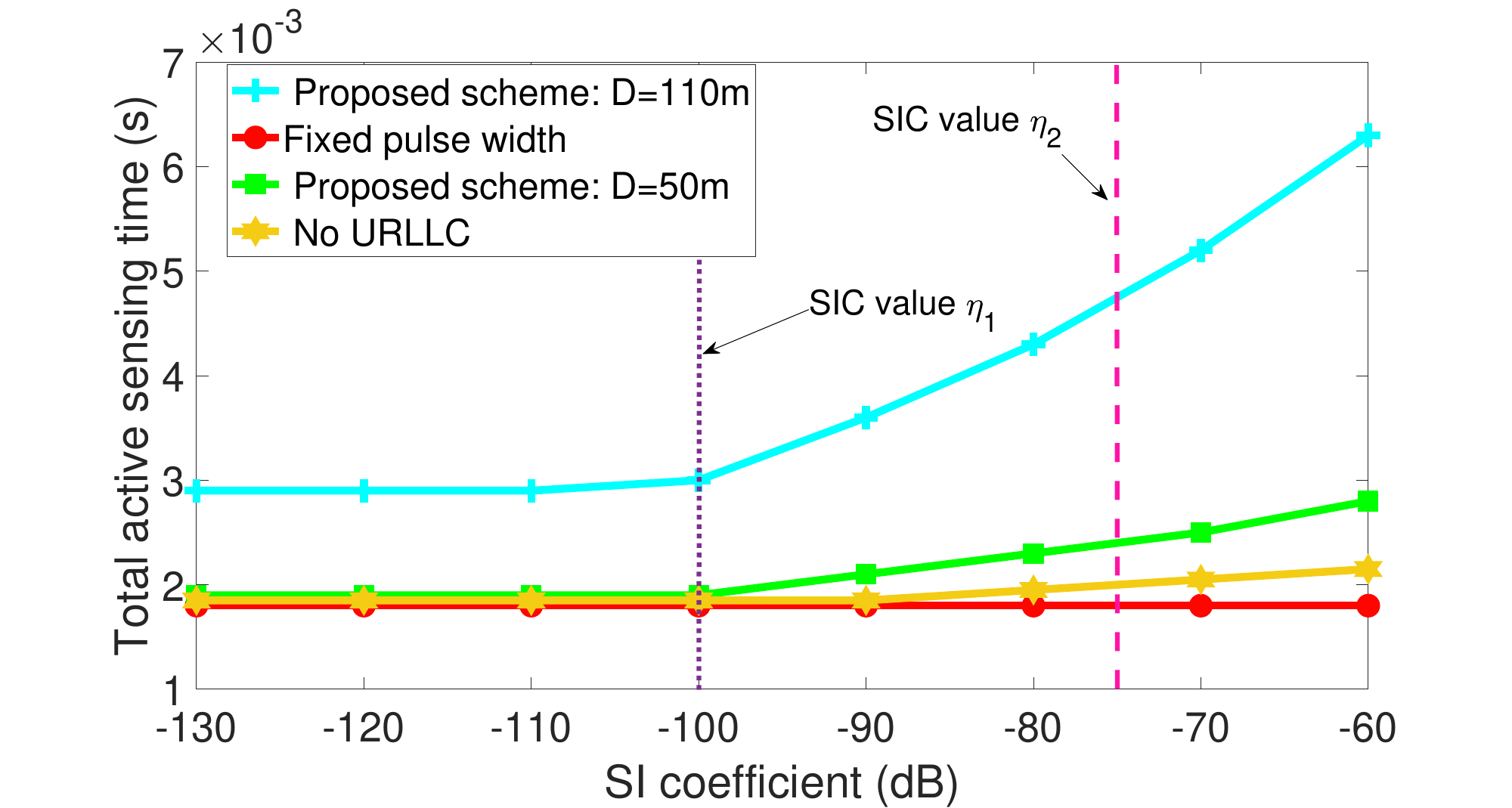}
}
\subfigure[Total energy consumption vs.\ $\eta$ for high accuracy sensing ($\zeta_{\text{max}}=0.01$) and low accuracy ($\zeta_{\text{max}}=0.05$).]{
\label{E_SI_b}
\includegraphics[width=0.475\textwidth]{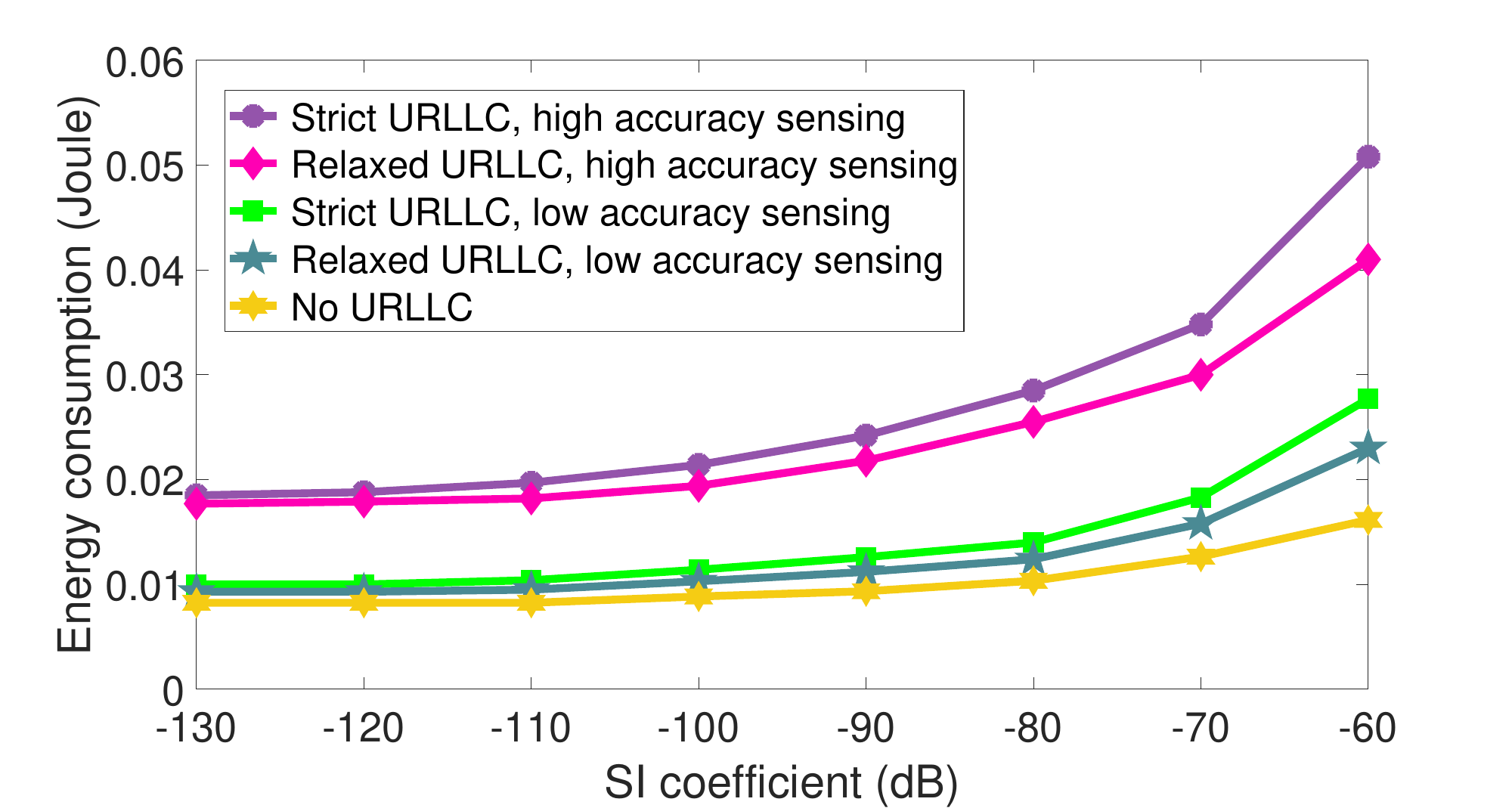}
}
\hfill
\subfigure[Total required active sensing time vs.\ $\eta$ for high accuracy sensing ($\zeta_{\text{max}}=0.01$) and low accuracy ($\zeta_{\text{max}}=0.05$).]{
\label{S_SI_b}
\includegraphics[width=0.475\textwidth]{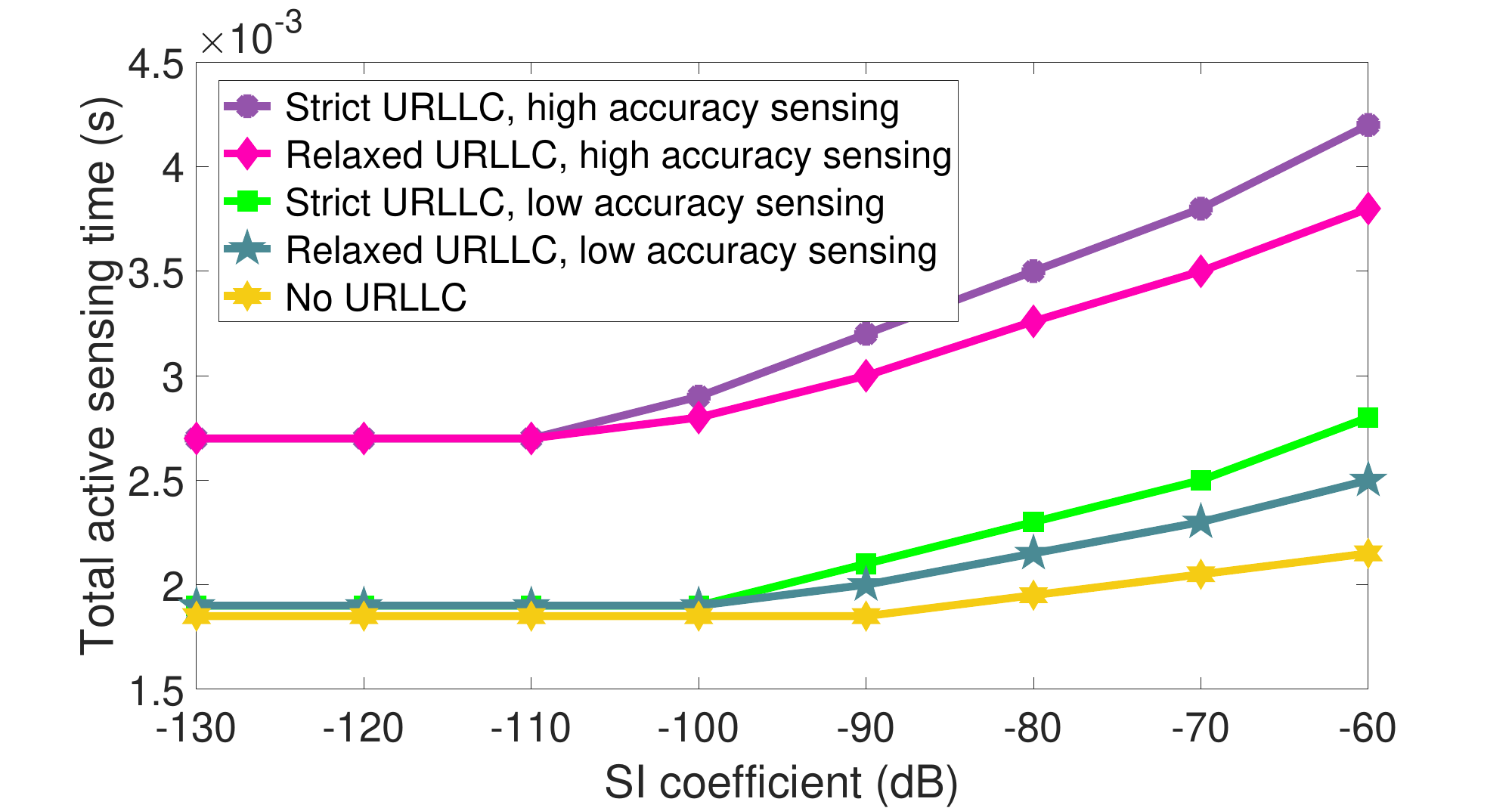}
}
\caption{Total energy consumption and total active sensing time of the DFRC-BS for $\zeta_{\text{max}}=0.05$, $D=50$m, $\epsilon_u=10^{-4}$, and $\nu_{\text{RCS}}=-10$dBsm.}
\label{S_SI}
    \vspace{-.8em}
\end{figure*}

\subsection{Scenario and Parameters}

We simulate a network with $K=3$ users and $E=3$ targets, positioned at ${(-D,0), (0,0), (D,0)}$, with the DRFC-BS located at $(0,D)$, where $D = 50 $m.
For URLLC traffic, we adopt the FTP3 traffic model, a widely used approach for characterizing URLLC traffic patterns  \cite{meredith2019study}.  
The detailed simulation parameters are presented in \cref{ISAC_parameters}, where the values of $\tilde{\lambda}_{k,n}$ and $\nu_{k,n}$ are chosen based on \cite{meredith2019study} to satisfy the per-user constraint. Furthermore, the time slot duration is assumed to be $\delta_t = 0.01\mathrm{s}$, following \cite{liu2020radarassisted}.
For comparison, we consider two baseline schemes as follows:
For baseline scheme 1, a fixed pulse width is assumed to evaluate the impact of sensing time optimization on overall network performance.
For baseline scheme 2, communication users are assumed not to follow URLLC regulations.
In this case, constraint $\text{C}$1 is omitted, and the Shannon capacity formula is used instead.

\begin{table}
\caption{System simulation parameters.}
\label{ISAC_parameters}
\centering
\begin{tabular}{lp{4.cm}l}
\toprule
Parameter & Description & Value \\
\midrule
$\sigma_{e}^{2}, \sigma_{k}^2$ & Noise power & $-80$ dBm \\
\midrule
$T$ & Time frame & $0.6$ s \\
\midrule
$\delta_{t}$ & Duration of one time slot & $0.01$ s \\
\midrule
$p^{\text{Rad}}_{\max}$ & Maximum power budget for sensing & $47$ dBm \\
\midrule
$p^{\text{Com}}_{\max}$ & Maximum power budget for communication & $30$ dBm \\
\midrule
$R_{\min}$ & Required  spectral efficiency & $0.5$ bits/s/Hz \\
\midrule
$\nu_{\text{RCS}}$ & Radar cross section & $-10$ dBsm \\
\midrule
$N_t$ & Number of transmit antennas & $8$ \\
\midrule
$N_r$ & Number of receive antennas at radar & $8$ \\
\midrule
$N_s$ & Number of sensing rounds & $300$ \\
\midrule
$t_{\min}$ & Minimum pulse width & $0.1~\mu$s \\
\midrule
$t_{\max}$ & Maximum pulse width & $4~\mu$s \\
\midrule
$\epsilon$ & Convergence tolerance & $10^{-3}$ \\
\midrule
$\epsilon_u$ & DEP threshold & $10^{-4}$ \\
\midrule
$\tilde{\lambda}_{k,n}$ & Arrived URLLC packets of UE $k$ & $250$ \\
\midrule
$\nu_{k,n}$ & Packet size per time slot & $32$ Byte \\
\midrule
$\zeta_{\max}$ & Maximum CRB threshold & $0.05$ \\
\midrule
$f_c$ & Carrier frequency & $1$ GHz \\
\midrule
$\lambda_1$ & Penalty factor & $10^{5}$ \\
\midrule
 & Rician factor & 5 dB\\
\bottomrule
\end{tabular}
\end{table}

\subsection{Impact of SI Coefficient $\eta$}

\Cref{E_SI_a,E_SI_b} depict the total energy consumption of the BS, encompassing both sensing and communication tasks as a function of $\eta$, the residual SI coefficient.
Additionally, \Cref{S_SI_a,S_SI_b} show the total accumulated sensing pulse width versus different values of $\eta$.
The total accumulated sensing pulse width represents the total active sensing time during which the BS transmits pulses across all sensing rounds and time slots.
For all the schemes shown in \Cref{E_SI_a,E_SI_b}, it is evident that as residual SI increases, both the required active sensing time and energy consumption gradually rise.
This trend becomes more pronounced for higher SI coefficients.
Specifically, when $\eta \geq -100$ \text{dB}, as shown in \cref{S_SI_a}, the total active sensing time sharply increases.
This indicates that the algorithm forces the system to extend the sensing duration to provide a sufficient observation window, partially mitigating the interference effects.
However, as shown in \cref{E_SI_a}, baseline scheme 1 (fixed pulse width) with $t_{p_n}= 0.1 \mu\text{s}$ requires significantly more energy to achieve the same sensing accuracy, particularly at higher SI levels.
This highlights the inefficiency of fixed sensing durations in handling residual SI.
In contrast, our proposed scheme, which incorporates sensing time optimization, achieves a tolerable energy demand while maintaining accurate target sensing, even in the presence of high residual SI.

As for baseline scheme 2, energy consumption and active sensing time in ISAC systems serving non-URLLC users are lower than in those supporting URLLC users, particularly at high levels of residual SI.
This outcome is expected, as URLLC users have stricter demands, including low latency and high reliability, which increase energy requirements.
The energy consumption gap between networks supporting URLLC and non-URLLC users reveals how high-quality communication affects sensing accuracy, especially for larger values of $\eta$.
The stringent requirements of URLLC traffic impact communication beamforming, leading to higher residual SI in the received echo signals.
Consequently, the system must consume more energy to maintain the same level of sensing accuracy for URLLC services. 
Additionally, the results confirm that greater distances between the radar and targets amplify energy and time consumption.
In these cases, the system's sensitivity to high residual SI levels is much greater compared to shorter distances.
For distant targets, high attenuation combined with elevated SI levels requires significant increases in both sensing time and energy to counteract these effects.
To provide context for SI rejection capabilities in practical environments, we reference two experimental studies.
\textcite{barneto2019full} demonstrate that up to $\eta_1=100$ dB of transmit-receive SI can be eliminated through a combination of passive cancellation, active RF cancelers, and digital suppression.
Similarly, \textcite{hassani2022joint} report approximately $\eta_2=75$ dB of analog SI cancellation using an adaptive filter. 

\Cref{E_SI_b} and \Cref{S_SI_b} further analyze the impact of sensing accuracy and reliability requirements on the system performance.
In these figure, we consider two sensing accuracy levels: 
\(\zeta_{\text{max}} = 0.01\) and \(\zeta_{\text{max}} = 0.05\), 
representing high and low accuracy, respectively. 
Increasing the sensing accuracy by five, with $\zeta_{\text{max}} = 0.01$ (compared to $\zeta_{\text{max}} = 0.05$), leads to significantly higher energy consumption due to the additional sensing observations required to reduce detection errors, as shown in \cref{E_SI_b}.
This emphasizes that ensuring highly accurate sensing is both challenging and energy-intensive.
Moreover, relaxed URLLC requirements ($\epsilon_u = 10^{-2}$) reduce resource and energy demands, as illustrated in  \Cref{E_SI_b,S_SI_b}.
On the other hand, stricter communication reliability requirements e.g., $\epsilon_u = 10^{-4}$ for URLLC use cases lead to higher energy consumption.

\begin{figure*}
\centering
\subfigure[Total energy consumption vs.\ parameter $D$.]{
\label{E_D_a}
\includegraphics[width=0.475\textwidth]{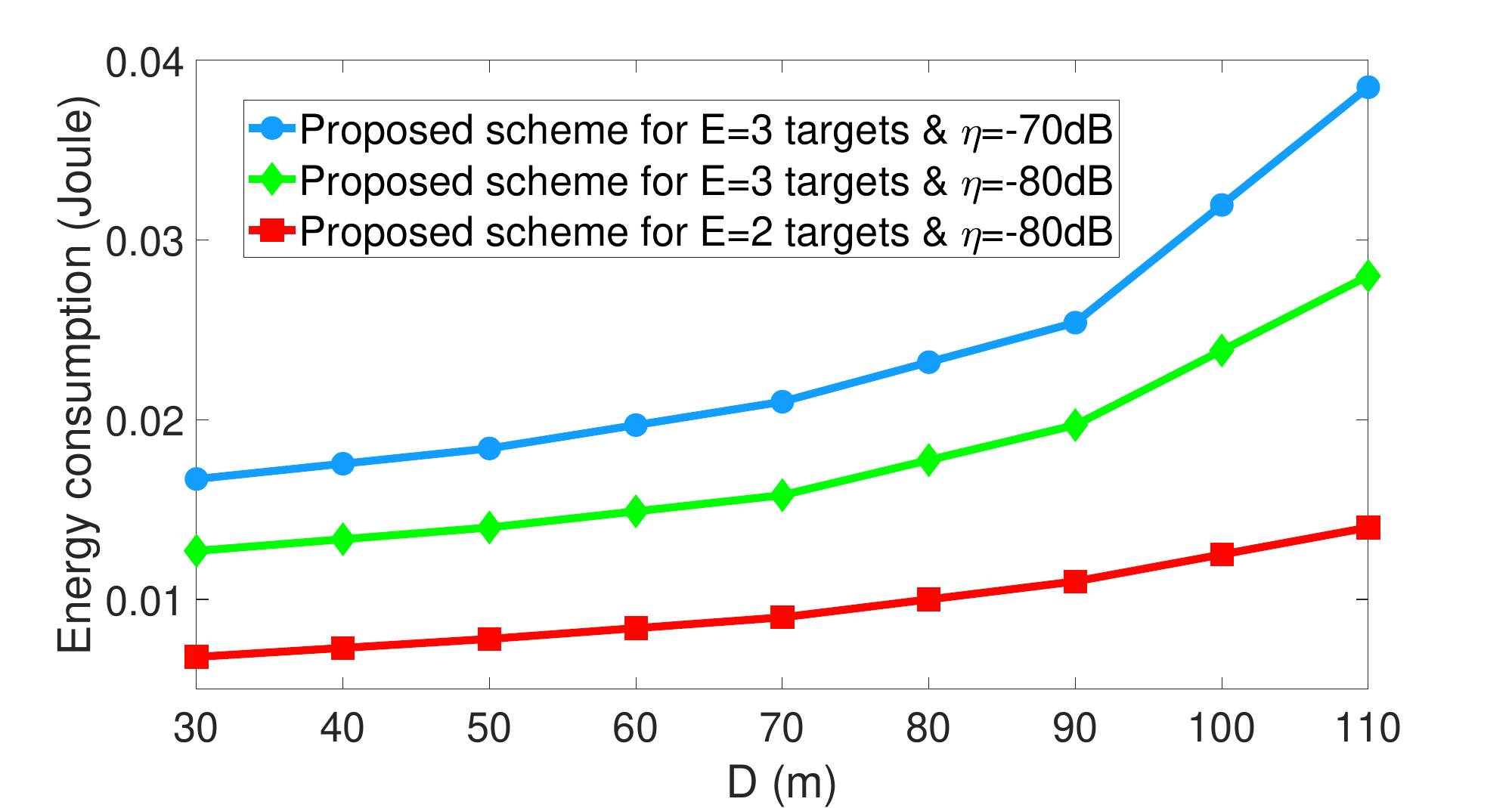}
}
\hfill
\subfigure[Total required active sensing time vs.\ parameter $D$.]{
\label{S_D_a}
\includegraphics[width=0.475\textwidth]{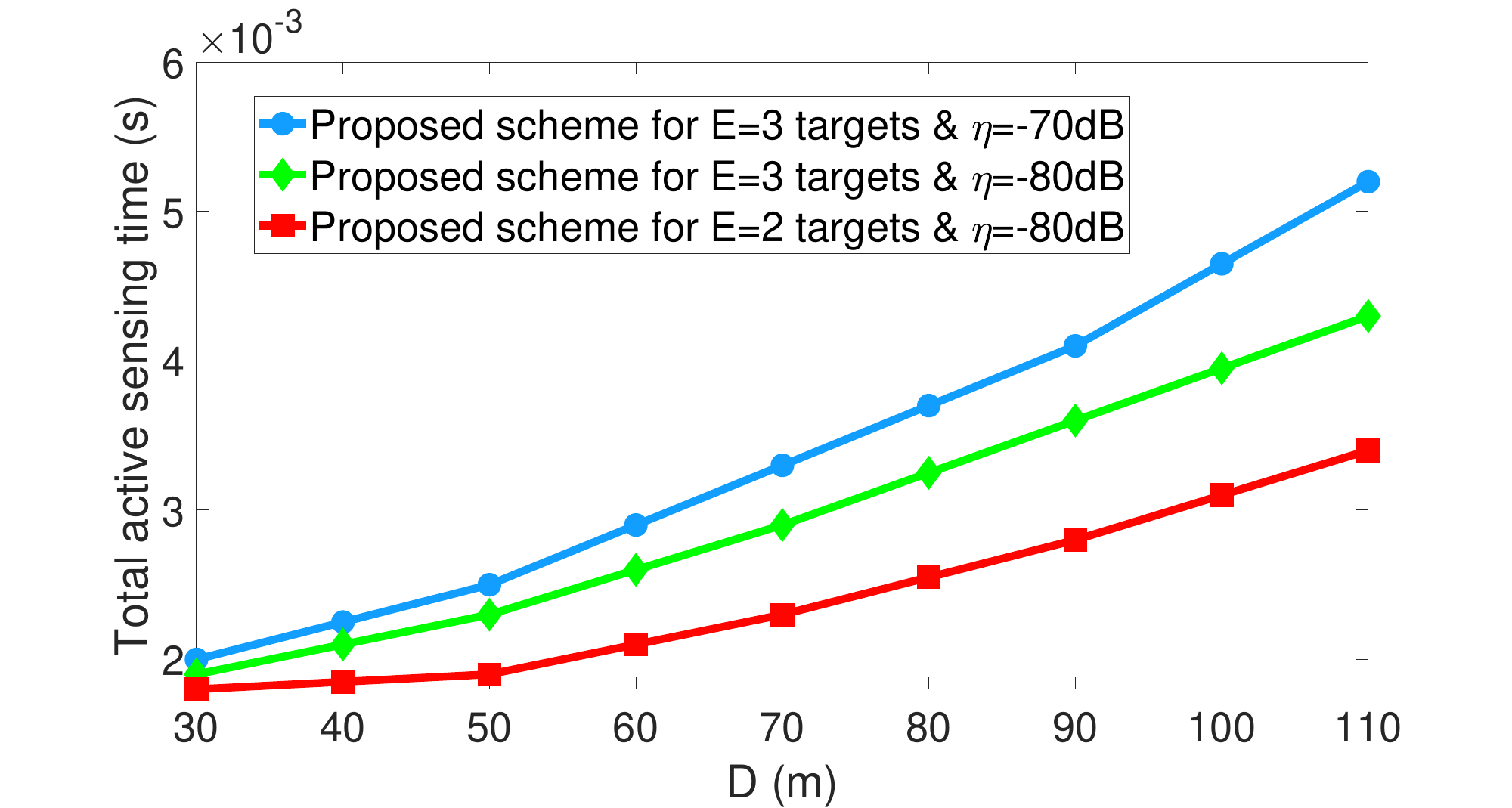}
}


\subfigure[Total energy consumption vs.\ $\nu_{\text{RCS}}$ for $D=110$m and $\eta=-80$dB.]{
\label{E_SI_a_RCS}
\includegraphics[width=0.475\textwidth]{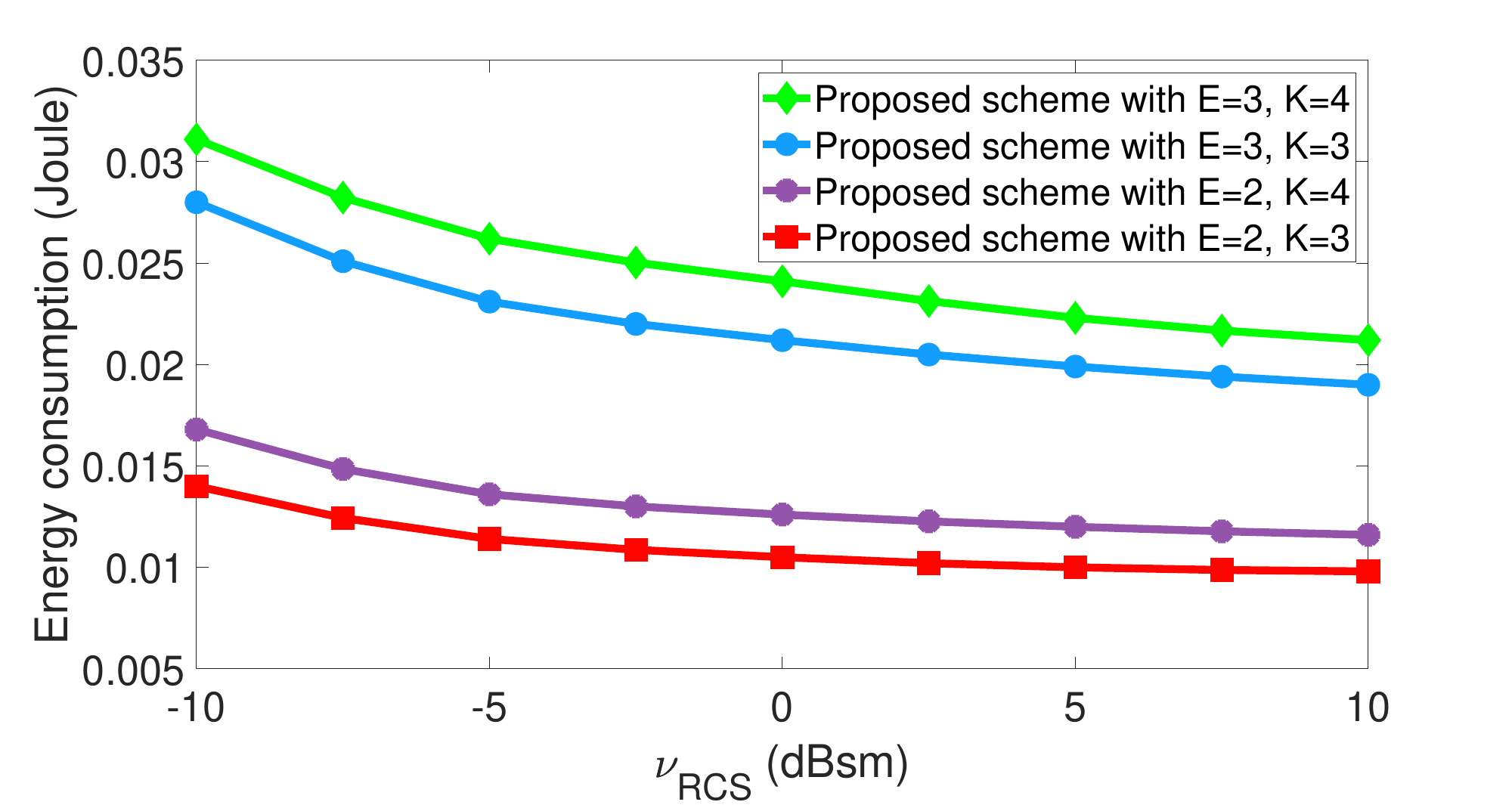}
}
\hfill
\subfigure[Total required active sensing time vs.\ $\nu_{\text{RCS}}$ for $D=110$m and $\eta=-80$dB.]{
\label{S_SI_a_RCS}
\includegraphics[width=0.475\textwidth]{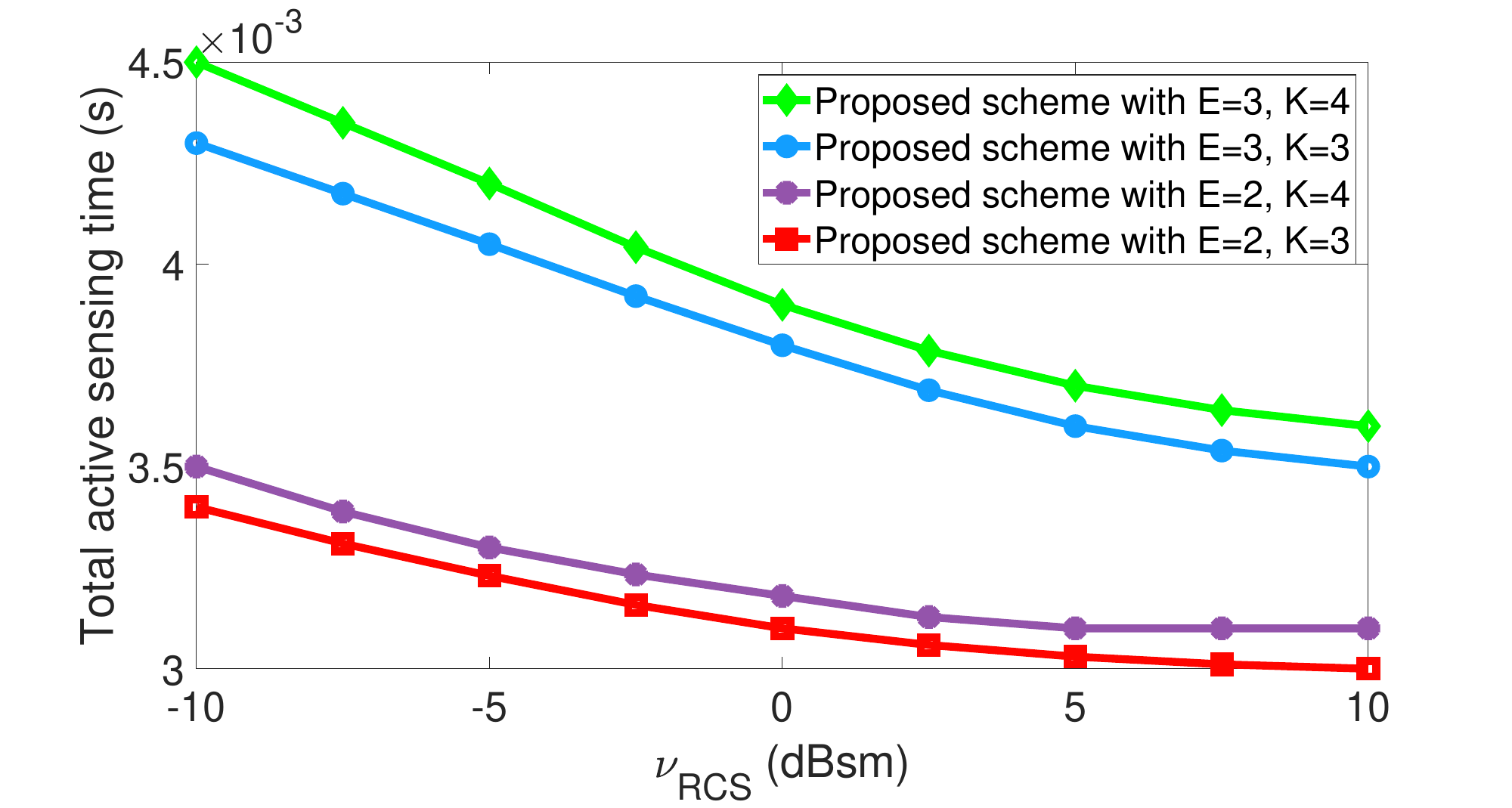}
}
\caption{Total energy consumption and total active sensing time of the DFRC-BS for $\zeta_{\text{max}}=0.05$, $D=50$m, $\epsilon_u=10^{-4}$, and $\nu_{\text{RCS}}=-10$dBsm.}
\label{S_SI_RCS}
    \vspace{-.8em}
\end{figure*}
\subsection{Impact of Distance $D$}

\Cref{E_D_a,S_D_a} show how increasing the distance from the BS, while still within its range, negatively affects sensing quality.
As the distance grows, the system compensates for the attenuation of echo signals by increasing both energy consumption and active sensing time for distant targets.
This effect becomes more pronounced when residual SI is significant, where target distance has a greater impact on system performance.
Additionally, the figures demonstrate how the number of targets affects active sensing time and energy consumption.
Reducing the number of targets from $E=3$ to $E=2$ reduces the required active sensing time and the resources needed for sensing.

\subsection{Impact of Radar Cross Section $\nu_{\text{RCS}}$}

The detectability of an object, determined by its RCS, denoted as $\nu_{\text{RCS}}$, depends on factors such as the target's material and size.
\Cref{E_SI_a_RCS} and \Cref{S_SI_a_RCS} analyze the radar's sensitivity to $\nu_{\text{RCS}}$ in terms of both energy consumption and active sensing time.
The results show that more detectable targets, which have higher 
$\nu_{\text{RCS}}$ values, require less energy and shorter pulse widths for sensing.
Additionally, the impact of increasing the number of communication users is considered. Adding users results in higher energy consumption and longer sensing pulse widths due to increased resource demands.
For example, when the number of users is increased to four, both energy consumption and active sensing time rise due to the greater potential for SI in the received echo signal, as well as the extra resources needed to support URLLC users in each time slot.


\section{Conclusion}

This paper proposes a novel framework for ISAC networks designed to support URLLC services while addressing the challenges of balancing sensing and communication tasks in resource-constrained environments.
By leveraging FD operation, the framework enables simultaneous transmission and reception, allowing the system to opportunistically utilize radar listening periods for communication tasks.
FD operation is crucial to mitigate SI through optimized beamforming and sensing time management, ensuring both efficiency and performance reliability.
The framework jointly optimizes key parameters, including sensing pulse duration and beamforming, to achieve accurate sensing and reliable communication while minimizing energy consumption.
Residual SI is identified as a significant factor influencing system performance, particularly in terms of energy consumption and sensing accuracy.
By incorporating sensing time optimization, the framework compensates for interference effects, effectively managing the trade-offs between sensing and communication tasks, even under strict QoS constraints for URLLC users.
Simulation results validate the proposed framework, demonstrating its ability to improve energy consumption while meeting the stringent requirements of URLLC services. 
The study highlights that supporting URLLC users incurs higher energy and resource demands compared to non-URLLC scenarios, underscoring the importance of intelligent resource allocation.
Moreover, the analysis evaluates the sensitivity of system performance to key factors such as residual SI levels, target distances, and sensing accuracy.
The findings reveal that increased target distances and higher residual SI levels significantly affect energy consumption and sensing time, while stricter URLLC and accuracy requirements further amplify resource demands.
This work offers a practical approach to addressing the challenges of FD-ISAC networks, providing insights into resource allocation and performance trade-offs.

\balance
\printbibliography

\end{document}